\documentclass[aps,prd,reprint,superscriptaddress,floatfix]{revtex4-2}

\usepackage{blindtext}
\usepackage[english]{babel}
\usepackage[utf8]{inputenc}

\usepackage[pdftex, pdftitle={Article}, pdfauthor={Author}]{hyperref} 
\usepackage{graphicx} 
\graphicspath{{}}						
									
\usepackage{amssymb}

\usepackage{amsmath}
\usepackage{tensor}
\usepackage{accents}

\usepackage{natbib}

\usepackage{empheq}
\usepackage{subcaption}
\usepackage[table]{xcolor}

\usepackage{booktabs}

\usepackage{xcolor}

\newcommand{\msun}{\ensuremath{M_\odot}}
\newcommand{\chimera}{{\sc Chimera }}
\newcommand{\thff}{\ensuremath{t_{\rm HFF}}}

\begin{document}

\title{Dependence of the Reconstructed Core-Collapse Supernova Gravitational Wave High-Frequency Feature on the Nuclear Equation of State, in Real Interferometric Data}

\date{2023}
\author{R. Daniel Murphy}
    \email[Correspondence email address: ]{rmurph16@utk.edu}
    \affiliation{Department of Physics and Astronomy, University of Tennessee, 1408 Circle Drive, Knoxville, Tennessee 37996-1200, USA}
\author{Alejandro Casallas-Lagos}
    \affiliation{Departamento de F\'isica, Universidad de Guadalajara, Guadalajara, Jal., 44430, M\'exico}
    \affiliation{Escuela de Ingeniería y Ciencias, Tecnológico de Monterrey, Monterrey, N.L., 64849, México}
\author{Anthony Mezzacappa}
   \affiliation{Department of Physics and Astronomy, University of Tennessee, 1408 Circle Drive, Knoxville, Tennessee 37996-1200, USA}
\author{Michele Zanolin}
    \affiliation{Department of Physics and Astronomy, Embry-Riddle Aeronautical University, Prescott, AZ 86301, USA}
\author{Ryan E. Landfield}
    \affiliation{National Center for Computational Sciences, Oak Ridge National Laboratory,P.O. Box 2008, Oak Ridge, Tennessee 37831-6164, USA}
\author{Eric J. Lentz}
    \affiliation{Department of Physics and Astronomy, University of Tennessee, 1408 Circle Drive, Knoxville, Tennessee 37996-1200, USA}
    \affiliation{Physics Division, Oak Ridge National Laboratory,
P.O. Box 2008, Oak Ridge, Tennessee 37831-6354, USA}
\author{Pedro Marronetti}
    \affiliation{Physics Division, National Science Foundation, Alexandria, Virginia 22314, USA}
\author{Javier~M.~Antelis}
\affiliation{Tecnologico de Monterrey, Escuela de Ingeniería y Ciencias, Monterrey, N.L., 64849, México}
\author{Claudia~Moreno}
\affiliation{Departamento de F\'isica, Universidad de Guadalajara, Guadalajara, Jal., 44430, M\'exico}

\date{\today} 

\begin{abstract}
    We present an analysis of gravitational wave (GW) predictions from five two-dimensional Core Collapse Supernova (CCSN) simulations that varied only in the Equation of State (EOS) implemented. The GW signals from these simulations are used to produce spectrograms in the absence of noise, and the emergent high-frequency feature (HFF) is found to differ quantitatively between simulations. Below 1 kHz, the HFF is well approximated by a first-order polynomial in time. The resulting slope was found to vary between 10--50\% from model to model. Further, using real interferometric noise we investigated the current capabilities of GW detectors to resolve these differences in HFF slope for a Galactic CCSN. We find that, for distances up to 1 kpc, current detectors have the ability to resolve HFF slopes differing by 4.4--15\%. For further Galactic distances, current detectors are capable of distinguishing the upper and lower bounds of the HFF slope for groupings of our models that varied in EOS. With the higher sensitivity of future GW detectors, and with improved analysis of the HFF, our ability to resolve properties of the HFF will improve for all Galactic distances. This study shows the potential of using the HFF of CCSN produced GWs to provide insight into the physical processes occurring deep within CCSN during collapse, and in particular its potential to further constrain the EOS through GW detection. 
\end{abstract}

\maketitle

\section{Introduction}

As gravitational-wave--based astronomy continues to mature with the detections of compact binary mergers \cite{GWTC2.1, GWTC3}, anticipation is growing for the advancements that will result from the detection of gravitational waves (GWs) from the next Galactic core-collapse supernova (CCSN). While LVK GW interferometers \cite{2015, 2015CQGra..32b4001A, Aso_2013} work to increase the ability to detect these events, it is the role of CCSN simulations to provide insight on what such detections can reveal about the Universe. This paper examines theoretical GW signals produced by CCSN simulations that utilized different nuclear equations of state (EOS) and uses current detection capabilities and analysis techniques to demonstrate the ability to resolve differences in GW signals due to different EOS. Our research builds on findings from \citet{PhysRevD.108.084027} and \citet{Lin_2023}, employing real LIGO interferometric noise from the second half of the O3b observing run \cite{KAGRA:2023pio}. This noise is analyzed to identify, reconstruct, and characterize the High Frequency Feature (HFF), observed in all CCSN GW signal predictions based on the two- and three-dimensional numerical simulations performed to date (for a review, see \citet{Abdikamalov_2021} and \citet{MezzZan_2024}).

CCSNe are stellar explosions that occur during the final life stage of massive stars -- \textit{i.e.} ${M_{\rm ZAMS}} \gtrsim 8$--10~\msun. The cores of these stars, having exhausted their nuclear fuel and dominated by degenerate, relativistic electron pressure, accumulate iron-group elements from silicon burning in the silicon layers encapsulating them, until they exceed the Chandrasekhar mass and collapse begins. The collapse of the inner core is halted when it surpasses nuclear saturation density and the strong nuclear force becomes repulsive. This causes the inner core to rebound outward -- \textit{i.e.} bounce -- and create a shock wave that proceeds to move through the supersonically infalling outer core. The inner, collapsed core continues to undergo deleptonization via neutrino losses, becoming a proto-neutron star (PNS) that will continue to evolve throughout the explosion. The shock moves outward and loses energy through nuclear dissociation of material passing through it and neutrino losses, which causes it to stall. As matter continues to fall onto the PNS behind the shock, all flavors of neutrinos are produced in the hot, shocked mantle of the PNS, and a region of net neutrino heating develops behind the shock. For most CCSNe, this neutrino heating mechanism plays a fundamental role in shock revival. See \cite{Muller_2016, Muller_2020, BurVar_2021} for reviews of CCSN theory and \cite{MezEndMes_2020} for a review of the neutrino heating mechanism in CCSNe, in particular.

From the moment of bounce onwards, each stage of a CCSN will generate  GWs, as described in \citet{Abdikamalov_2021} and \citet{MezzZan_2024}. The present study focuses on the high-frequency GW signal arising from accretion onto the PNS, as well as sustained Ledoux convection due to lepton gradients deep within the PNS. These lepton gradients become large enough to cause Ledoux convection in the PNS around 100 ms postbounce and are sustained for on the order of seconds as neutrinos continue to diffuse out of the PNS. During this time, the PNS contracts. This contraction causes the frequency of the GWs produced by Ledoux convection to increase with time until the PNS reaches a stable radius, at which point the GW frequency remains constant while Ledoux convection continues. We call this feature the High Frequency Feature (HFF), using nomenclature introduced in \citet{PhysRevD.108.084027}. Based on analysis of the quasi-normal modes of oscillation of the PNS, previous studies have concluded that the HFF corresponds to the frequency evolution of either a $g$-mode oscillation \cite{Marek_2009, Cerda-Duran_2013, Kuroda_2017, Muller_2013, Kawahara_2018} or a combination of $g$-mode and $f$-mode oscillations \cite{Sotani_2016, Morozova_2018, Torres-Forne_2017,Torres-Forne_2019, Radice_2019, SotTak_2020, Sotani_2021, Zha_2024}. The classification of these modes may connect to the PNS properties, but such analysis is beyond the scope of this paper.

In CCSN simulations, the EOS not only connects the state and thermodynamic variables, but also regulates the radius and contraction of the PNS. An EOS is called `stiff' if it has more resistance to compression and thus yields a larger NS radius, and `soft' if it is more easily compressed, yielding a smaller NS radius. Different EOS result in different dynamical evolution of the PNS and influence GW generation from bounce and for several seconds afterwards as the PNS evolves. A clear goal of GW analysis, therefore, is to quantify the differences in GW signals that arise from different EOS. In addition to the HFF, prominent GW features often studied in connection to the EOS in CCSN simulations are the Standing Accretion Shock Instability (SASI) \cite{Marek_2009, Kuroda_2016, Kuroda_2017, Kotake_2017, Powell_2021, Kuroda_2022}, which generates GWs at lower frequencies, and the GW burst and subsequent ring down caused by core bounce for models with rotation \cite{Dimmelmeier_2008, Yasutake_2007, Andersen_2021, Richers_2017, Afle_Brown_2021, Edwards_2021, Mitra_2024}.

Early connections between GW signals in CCSNe and the EOS were noted in \citet{Seidel_1988}. Using an EOS that parameterized the stiffness of the EOS in spherically symetric simulations, they found that stiffer parameterizations led to slightly more efficient GW production. \citet{Marek_2009} investigated the GW and neutrino emissions in two-dimensional CCSN numerical simulations using a 15~\msun\ progenitor, comparing stiff and soft EOS. They observed that PNS formed under the soft EOS, being more compact, generated GWs of higher amplitude than those produced under the stiff EOS, after 100 ms post-bounce. Additionally, they noted that (i) the peak frequency of the waveform spectrum for the HFF was above 300 Hz, (ii) the peak frequency for the SASI was below 200 Hz, with (iii) both frequencies being elevated in the case of the softer EOS.

\citet{Andersen_2021} used EOS models based on a Skyrme-type force to investigate the sensitivity of the effective nucleon mass and isoscalar incompressibility modulus on the GW signal in two-dimensional CCSN simulations. They found that the peak of the GW frequency spectrum $f_{\rm peak}$ increases with effective nucleon mass parameter, while no variations seemed to occur in the GW signal from variation of the isoscalar incompressibility modulus parameter. This effect persisted in simulations with central angular speeds of 1 rad $\mathrm{s}^{-1}$, though the GW amplitudes decreased, and the effect was not discernible at central angular speeds of 2 rad $\mathrm{s}^{-1}$. Additionally, this study produced second-order polynomial fits to the HFF and reported different coefficients across the nucleon effective mass parameter.

\citet{Morozova_2018} compared calculated PNS oscillation frequencies directly with GW signals produced from two-dimensional CCSN simulations of 10, 13, and 19 \msun\ progenitors. Quantitative differences were reported for GW amplitudes, GW energy emitted, and for the frequencies of PNS quasi--normal-mode oscillations across the three EOS they considered and for each progenitor mass. At later times, the $f$-mode oscillations matched well with the peak frequency evolution of the GW signal, and a quadratic polynomial fit using the $f$-mode oscillation frequencies resulted in distinct coefficients for each EOS.

\citet{Camelio_2017} investigated the oscillation modes of the PNS using three different EOS, and found that the time it took the PNS to reach the maximum (minimum) oscillation frequency of the $f$-($g$-)mode varied by as much as 0.5 seconds. Using  data from one-, two-, and three-dimensional CCSNe simulation data, \citet{SotTak_2020} computed PNS oscillation frequencies and found only minor differences in their evolution using the DD2 and TGTF EOS. By analyzing the $f$-mode PNS oscillations together with spacetime oscillations, $w$-modes, \citet{Sotani_2019} determined the different mass and radius evolution of the PNS for a CCSN model using a 15 \msun progenitor and both the SFHx and TM1 EOS, in the absence of noise.

Universal relations describing the GW peak frequency evolution as either an $f$- or $g$-mode were investigated in \cite{Torres-Forne_2019, Sotani_2021, Wolfe_2023}. \citet{Bizouard_2021, Bruel_2023} investigated parameter estimation using universal relations in Gaussian simulated noise. In each study, different EOS were considered, but the focus was the accuracy of the parameterization independent of the EOS. A similar modal analysis and universal relation study for the data presented here could be performed but is beyond the scope of this paper.

\citet{Yasutake_2007} studied GW signals from CCSN simulations using a phenomenological EOS that allowed for a first-order quantum chromodynamics (QCD) phase transition. In the absence of rotation, they saw a ~10\% increase in GW amplitude immediately after bounce when compared to the same EOS without a phase transition. \citet{Zha_2020} combined the STOS EOS with the MIT bag-model EOS to study similar QCD phase transitions in CCSNe. They found that the core experienced a secondary collapse at the QCD phase transition that resulted in GW amplitudes ~30 times stronger than typical signals from purely hadronic EOS.

For failed CCSNe, \citet{Pan_2018} investigated the GW signals from black-hole--producing two-dimensional CCSN simulations starting from a 40~\msun\ progenitor, using four different EOS. They noted visible differences in the characteristic strain evolution $h_{\rm char}$, and they also noted that black-hole--formation times varied from 450 ms to 1300 ms across the EOS used. In the context of pulsational pair instability supernovae, \citet{Powell_2021} investigated GW signals in Gaussian noise for current and future detectors for 85 and 100 \msun progenitors, using the LS220, SFHo, and SFHx EOS. They find that the HFF increases slowest for SFHx and fastest for LS220.

\citet{Jakobus_2023} examined the GW signals from two-dimensional general relativistic CCSN simulations strating from 35 \msun and 85 \msun progenitors, using the SFHx EOS as well as a chiral mean field EOS with a smooth crossover to quark matter. The CMF EOS causes the HFF to increase slightly faster than the SFHx EOS, for both progenitors. They also identified a low-frequency feature that decreases in frequency with time. In the SFHx models, this feature only appeared in the 85 \msun progenitor, but it was of much higher frequency and decreased more slowly than in the CMF EOS models. This feature emanated from deep within the core of the PNS, allowing for its use as a probe to study nuclear matter beyond nuclear saturation density.

Beyond the effects of EOS, the progenitor properties can affect the HFF as well. Rotation of the progenitor tends to have a stabilizing effect on the radius of the PNS, leading to a slower contraction time. This would tend to decrease the slope of the HFF as seen, \textit{e.g.}, in \cite{Andresen_2019, Pajkos_2019, Pajkos_2021, Pan_2021, Takiwaki_2021}. The mass of the progenitor will affect the amplitude of the GW strains, as well as the slope of the HFF \cite{Andresen_2017, Vartanyan_2019, Mezz_D_GW_2023}, though a systematic study is necessary to quantify the effect more precisely. Additionally, \citet{Wang_2024} and \citet{Jardine_2022} show that the internal structure of the progenitor and the presence of strong magnetic fields can affect the HFF slope, respectively. In this study, we choose to study the effects of the EOS on the HFF in isolation of all progenitor parameters. In essence, we are assessing the maximum amount of information that could be gathered about the EOS, at least in our approach, in the event of a Galactic core collapse supernova gravitational wave detection, assuming that all of the other parameters affecting the HFF slope are known.

Previously, \citet{PhysRevD.108.084027} formulated a machine learning (ML) approach using real LIGO interferometric noise to estimate the slope of the HFF peak over time across different two- and three-dimensional CCSN models incorporating distinct EOS. The results indicated that the HFF slope is influenced by factors such as the progenitor mass, rotation, metalicity, and EOS. In scenarios with non-rotating progenitors and consistent EOS across CCSN simulations, the HFF slope without noise ranged from 3406, 1907, to 1288 Hz $\mathrm{s}^{-1}$ for progenitors of 15, 20, and 35 \msun, respectively, demonstrating variabilion in HFF slopes up to 63\% for a mass change of 57\%. For a 15~\msun\ progenitor rotating at 0.5 rad $\mathrm{s}^{-1}$, the HFF slope found was 2246 Hz $\mathrm{s}^{-1}$, reflecting a 34\% reduction from its non-rotating counterpart of the same mass. These signals were introduced into real interferometric LIGO noise through a cWB event production analysis, in standard configuration, for different Galactic distances 1, 5, and 10 kpc. The ML algorithm managed to predict the slope with an error under 10\% for nearly all models except three. Moreover, the orientation of the source was found to have a minimal impact on the slope estimation of the HFF, despite variations in the intensity of CCSN GW signals with orientation, as discussed in \citet{Muller_2016, Morozova_2018}. Differences in HFF slope due to orientation angle with the GW source were under 5\%. The study confirmed the reliable estimation of the HFF slope in real interferometric noise and highlighted several physical parameters that influence this estimation.

This paper continues the development of the HFF as a tool to gain physical insight from Galactic-CCSN-produced GW signals and, in particular, examines the effects of the EOS on it. We extract GW signals from two-dimensional, axisymmetric CCSN simulations \cite{Landfield_thesis,Landfield_EOS_paper} performed with the \chimera\ code \cite{Bruenn_2020}. Five models were considered that employed five distinct EOS in the CCSN simulations: DD2, FSUGold, IUFSU, SFHo, and SFHx. These EOS were chosen for their compatibility with current experimental and observational constraints, as discussed in \citet{Tews_2017}. We note, these EOS do not cover the full range of permissible EOS. Discussions on the explosion characteristics of the two-dimensional simulations can be found in \citet{Landfield_thesis}. These GW signals are first analyzed in the absence of noise to determine the intrinsic differences in HFF slope due to differing EOS in the models considered and signal processing settings. Using these intrinsic slope values, we then investigate our ability to resolve different EOS through the detection of GWs and analysis of the HFF using real interferometric LIGO data in a Coherent-Waveburst (cWB)-event production analysis \cite{Klimenko_2016}. Our results spans Galactic distances of 1, 5, and 10 kpc in an equatorial orientation, but varying orientations are achievable by adjusting the term $1/r$ equivalently to a multiplication by the cosine of the orientation angle $\theta$, as detailed in \cite{Creighton:2011zz}. 

The paper is organized as follows: Section \ref{sec:Models_methods} outlines the differences between each EOS and provides the theoretical framework for extracting GW signals from \chimera\ data. Section \ref{sec:GW_A} analyzes the HFF in the absence of noise. Section \ref{Sec:HFF_Estimation} describes the methods of detection and HFF slope estimation using real LIGO noise for GW signals at 1, 5, and 10 kpc as well as rescalings of the results for future configurations. Section \ref{Sec:Disc} discusses these results in the context of past GW EOS studies and the capabilities of current and next generation detectors. Section \ref{Sec:conc} summarizes our findings and outlines potential future work.

\section{Models and Methods in EOS numerical simulations}\label{sec:Models_methods}

\subsection{\chimera\ Simulations}\label{sec:Chimera}

All CCSN simulations analyzed in this paper were computed with the neutrino-radiation hydrodynamics code \chimera\cite{Bruenn_2020}.
\chimera\ simulates CCSNe using Newtonian self-gravity with a general relativistic spherical, monopole correction, Newtonian hydrodynamics, multigroup flux-limited diffusion neutrino transport in the ray-by-ray approximation, and a nuclear reaction network. Neutrino-matter interactions in \chimera\ include electron capture on protons and nuclei, electron--positron annihilation, and nucleon--nucleon bremsstrahlung, along with their inverse weak reactions. Neutrino scattering processes included are isoenergetic scattering on nuclei, neutrino-electron scattering, and neutrino-nucleon scattering.

The two-dimensional simulations used in this paper are from the E-series examination of nuclear EOS effects \cite{Landfield_thesis,Landfield_EOS_paper}.
The simulations are referred to by the EOS label such that the simulation implementing the DD2 EOS will be referred to as E-DD2.
The initial conditions for each simulation are that of the 15~\msun\ pre-supernova progenitor of \citet{Woosley_2007}, which is a non-rotating progenitor of Solar metallicity. The same 15~\msun\ progenitor is used in the three-dimensional simulation in \citet{Mezz_D_GW_2023} from which we compare two- and three-dimensional results. For the two-dimensional simulations, there are 720 adaptive radial zones and 240 angular zones of fixed angle. The radial zones redistribute with time to sufficiently resolve the shock and the PNS surface.

\subsection{Included Equations of State}
Each of the EOS used in the E-series uses a relativistic mean field (RMF) approximation to model nuclear interactions. They differ in the method of parameterization of the RMF, and in how the values for those parameters are determined. Each of these EOS extend nuclear statistical equilibrium (NSE) models, applicable from low densities up to a few tenths of nuclear saturation density, to RMF-based models to capture the high-density behavior of the EOS. In NSE, the ensemble of nuclei and non-uniform nuclear matter are represented via a statistical model as described in \citet{Hempel_2010}. Nuclei are treated as Maxwell-Boltzmann particles, and the nucleons are treated with RMF models. This representation includes several thousand nuclei with binding energies either taken from experimental measurements or theoretical nuclear structure calculations. Where nuclear structure calculations are used, they are calculated using the same RMF parameterization as the one applied to nucleons, if available. 

The DD2 EOS uses a density-dependent (DD) RMF for nuclear interactions \cite{Typel_2010} that come from describing the meson scalar and vector self energies with density-dependent forms. The ``2'' differentiates it from the original DD EOS \cite{Typel_2005}. The DD EOS used theoretical nucleon masses instead of experimental values as is done for the DD2 EOS.

The FSUGold EOS implements a RMF treatment with additional $\omega$- and $\rho$-meson couplings in the Lagrangian \cite{Todd-Rutel_2005}. This results in softening the EOS of symmetric nuclear matter and modifies the density dependence of the symmetry energy.

The IUFSU EOS builds upon the FSUGold EOS with more recent experimental and observational data \cite{Fattoyev_2010}. This introduces two new empirical parameters that affect the EOS at different densities. At intermediate densities, the EOS is softened by reducing the neutron skin thickness of ${}^{208}$Pb. At high densities, the EOS is stiffened by increasing the maximum neutron star mass relative to FSU predictions.

SFHo and SFHx are EOS that use a RMF parameterization that was uniquely fitted using observational neutron star radius determinations \citet{Steiner_2013}. Each EOS is based on the paramterization of \cite{Steiner_2005}, with parameters varied such that saturation properties agree with predictions from nuclear masses, giant monopole resonances, and so that the maximum neutron star mass is larger than 1.93~\msun. Additionally, the ordinary model, SFHo, uses the most probable neutron star mass-radius curve from \citet{Steiner_2010} to further constrain the parameters of the RMF, while the extreme model, SFHx, attempts to minimize the radius of low-mass neutron stars in its parameterization. As a result of this, the logarithmic derivative of the symmetry energy for SFHx is in the lower part of the acceptable range, 20-120 MeV.

In addition to these five EOS, the E-series also includes a simulation that used the Lattimer-Swesty EOS \cite{LS_1991} with a nuclear incompressibiliy of $K=220$ MeV combined with the BCK EOS \cite{BCK_1985} at densities below $10^{11}$ g cm$^{-3}$, referred to collectively as LSBCK. This EOS was not investigated in terms of resolvability because it is no longer within the allowable region of EOS, as shown in \cite{Tews_2017}. However, this is the same EOS that was used for the three-dimensional simulations in \citet{Mezz_D_GW_2023}. This allows for a separate, but necessary, analysis of the effect of dimensionality on HFF properties, in Section \ref{sec:2d3d}.

\subsection{Gravitational Waves in CCSN Simulations}\label{sec:GW_Extract}

The procedure for extracting gravitational wave strains is described in full in \citet{Mezz_D_GW_2023}. The quadrupole moment, as the lowest order surviving multipole, is the primary source of gravitational waves, and so the gravitational wave signal is approximated as arising solely from this term. In the weak field approximation, the quadrupole moment of the transverse-traceless gravitational wave strain is given by
\begin{equation}
h^{TT}_{ij}=\frac{G}{c^4}\frac{1}{r}\sum^{+2}_{m=-2}\frac{d^2I_{2m}}{dt^2}\left(t-\frac{r}{c}\right)f^{2m}_{ij}, 
\end{equation}
where $i$ and $j$ span the spherical coordinates $r$, $\theta$, and $\phi$ and $f^{2m}_{ij}$ are the tensor spherical harmonics. The mass quadrupole is 
\begin{equation}
    I_{2m}=\frac{16\sqrt{3}\pi}{15}\int\tau_{00}Y^*_{2m}r^2dV,
\end{equation}
with $dV=r^2\sin\theta dr d\theta d\phi$ and $\tau_{00}$ being the rest mass density, $\rho$, in the weak field limit. The amplitude of the gravitational wave is then defined by
\begin{equation}
    A_{2m}\equiv\frac{G}{c^4}\frac{d^2I_{2m}}{dt^2}. \label{eq: A_2m}
\end{equation}

In the two-dimensional case considered here, there is only one non-zero polarization of gravitational wave strains, namely
\begin{equation}
    h_+=\frac{h^{TT}_{\theta\theta}}{r^2}.
\end{equation}

To investigate the HFF, we produce spectrograms without detector noise from the gravitational wave strains through the use of the short-time Fourier transform (STFT) given by
\begin{equation}
S(\tau,f)=\int_{-\infty}^{\infty}h_+(t)w(t-\tau)e^{-i2\pi ft}dt,
\end{equation}
with $w$ being the analysis window of finite duration that captures the signal power centered around time $\tau$ and frequency $f$. The spectrogram is then given by
\begin{equation}
P(\tau,f)=\|S(\tau,f)\|^2.
\end{equation}
For the spectrograms produced in Section \ref{sec:GW_A}, we used a Kaiser windowing function given by
\begin{equation}
    w(n)=\frac{I_0\left(\beta\sqrt{1-\left(\frac{2n}{N}-1\right)^2}\right)}{I_0(\beta)},
\end{equation}
for a window containing $N-1$ points, where $I_0$ is the zeroth-order modified Bessel function of the first kind. The shape factor, $\beta$, is a parameter that changes the equivalent noise bandwidth of the signal between the maximal case ($\beta=0$) of a rectangular window and the minimal case ($\beta=40$). Here, noise is introduced through processing the signal with the windowing function, and $\beta$ is a way to mitigate the effect of this processing noise without completely changing the windowing scheme. 
\section{Slope Estimation Without Noise}\label{sec:GW_A}
For the spectrograms in this section, the GW signal was sampled at 5000 Hz. For the windowing function, we chose a shape factor of $\beta=6$, which approximates a Hann windowing scheme, with 45 millisecond window segments overlapping adjoining segments by $93.\bar{3}\%$. This resulted in spectrograms that were visually well resolved in both the time and frequency domains.

    \begin{figure*}[!ht]
        \centering
            \includegraphics[width=8.6cm]{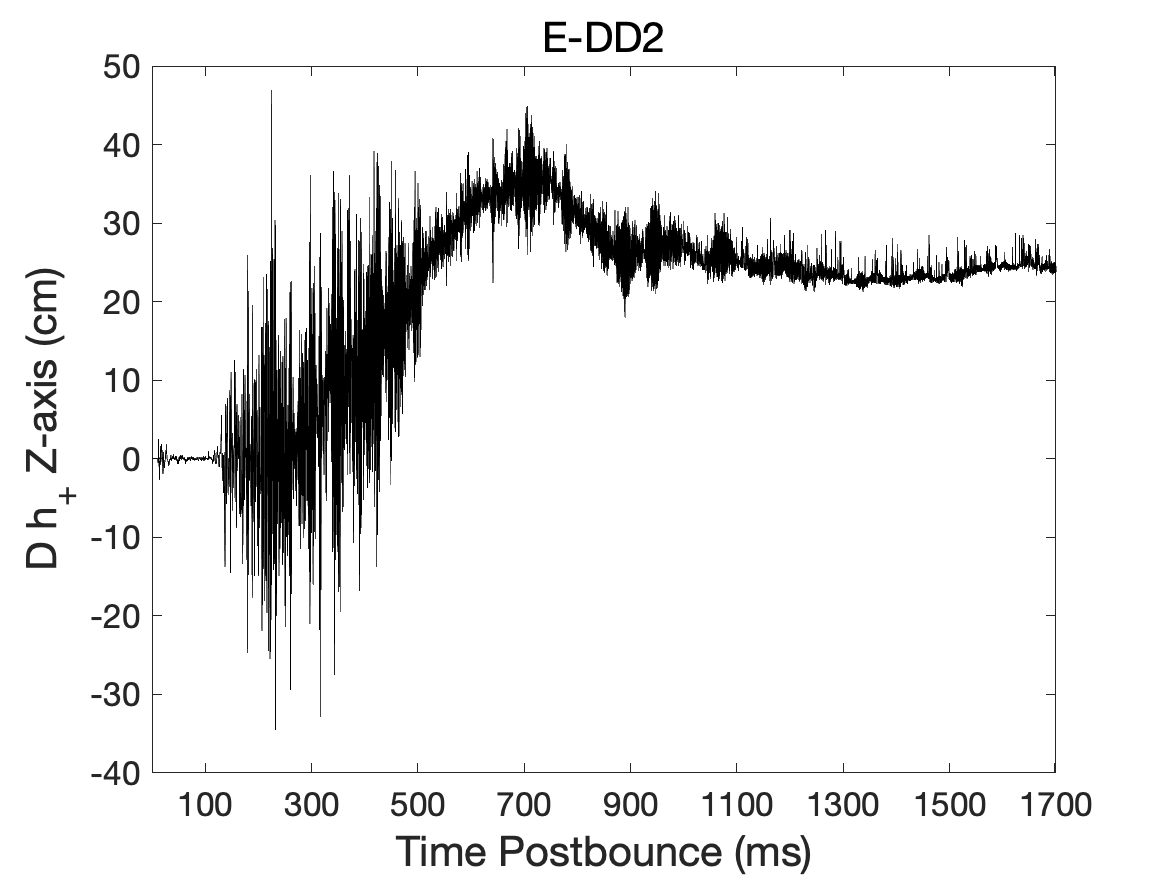}
        \hfill
            \includegraphics[width=8.6cm]{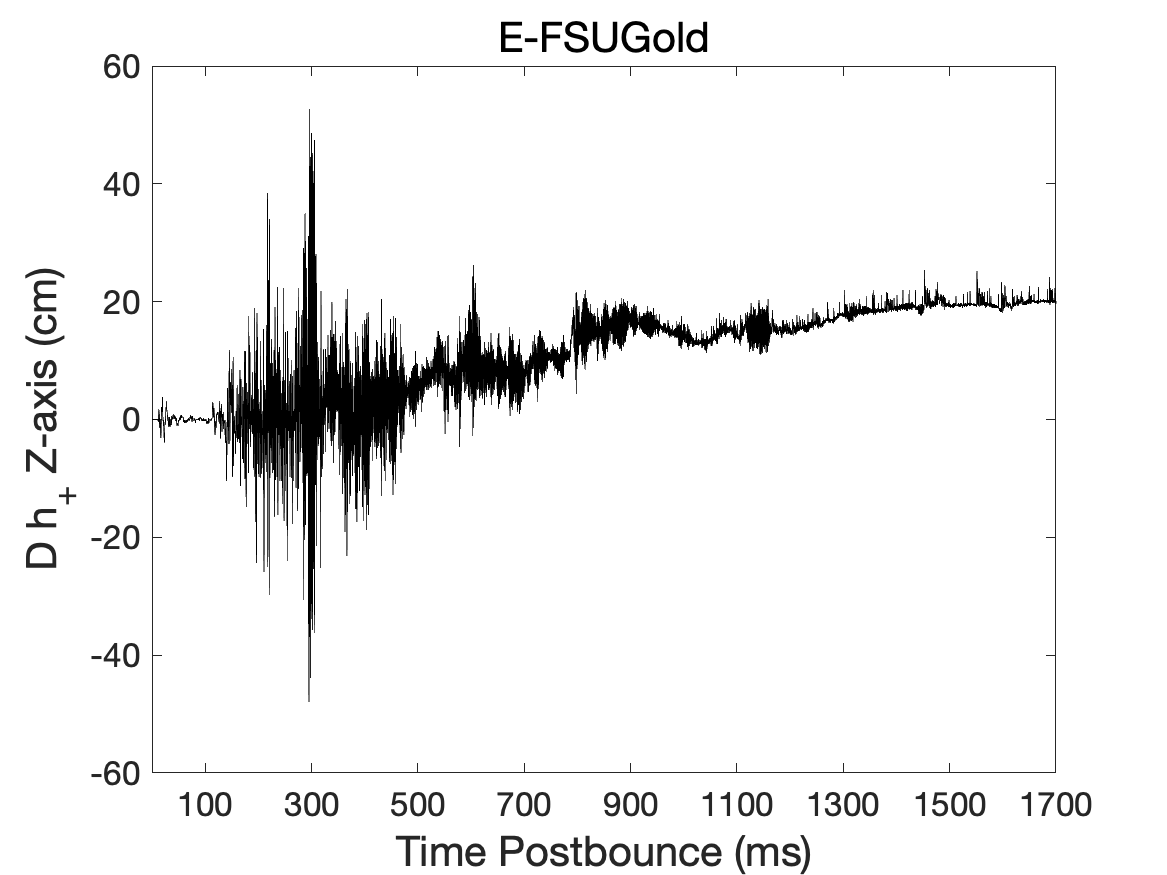}
        \hfill
        \vspace{4mm}
            \includegraphics[width=8.6cm]{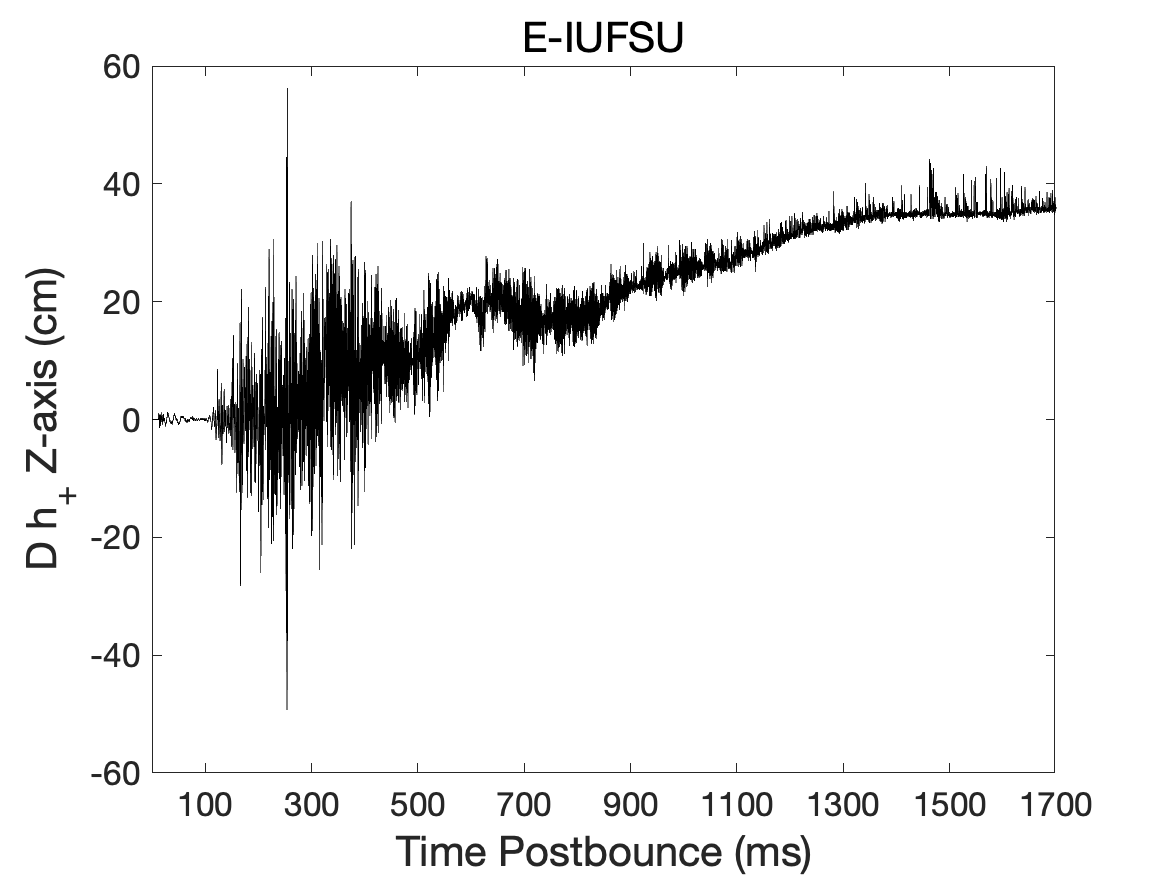}
        \hfill
            \includegraphics[width=8.6cm]{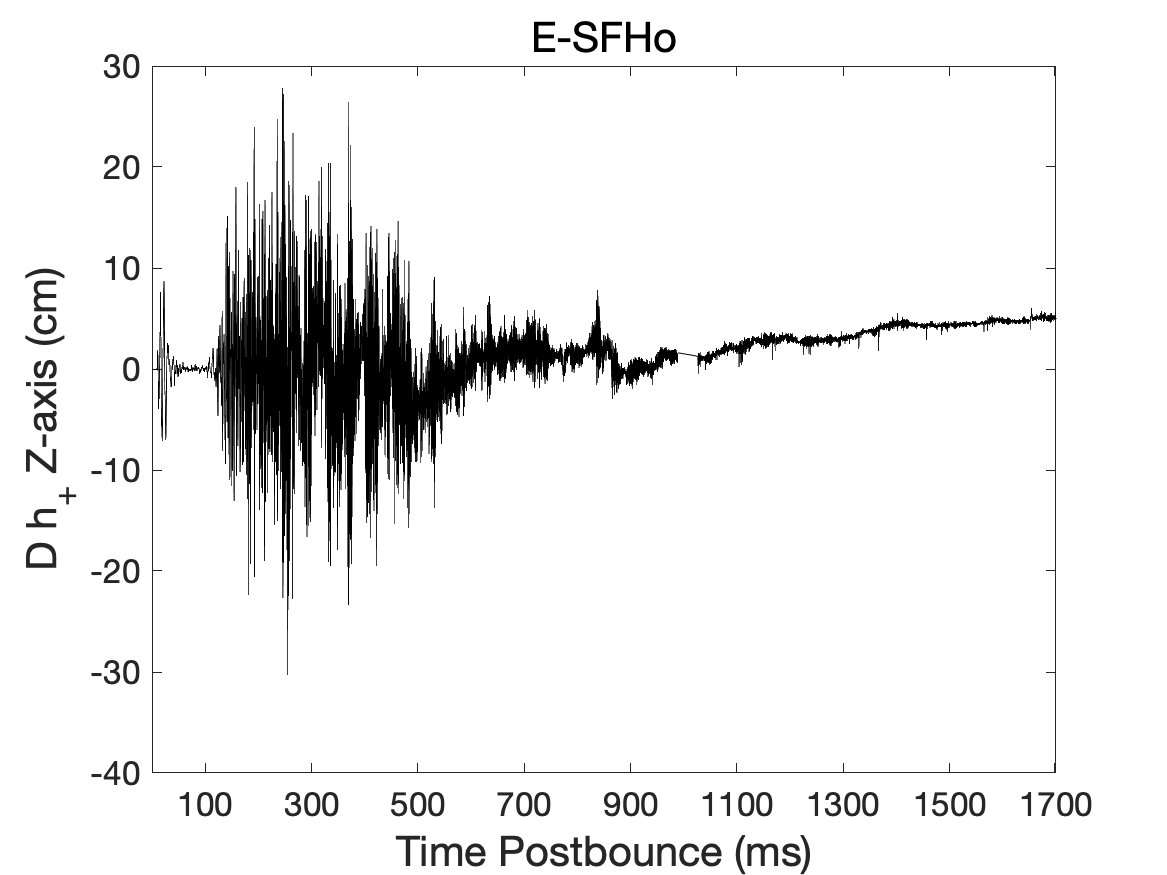}
        \hfill
        \vspace{4mm}
            \includegraphics[width=8.6cm]{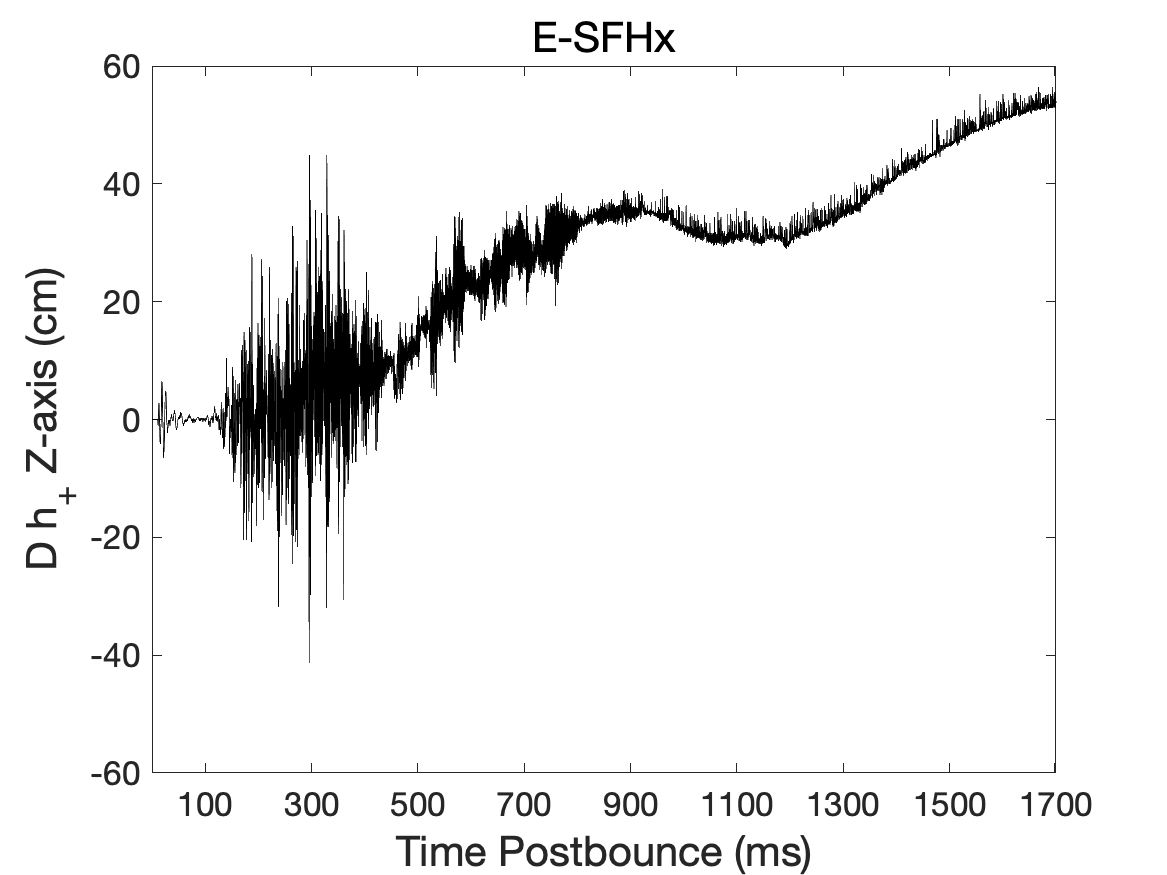}
        \hfill
            \includegraphics[width=8.6cm]{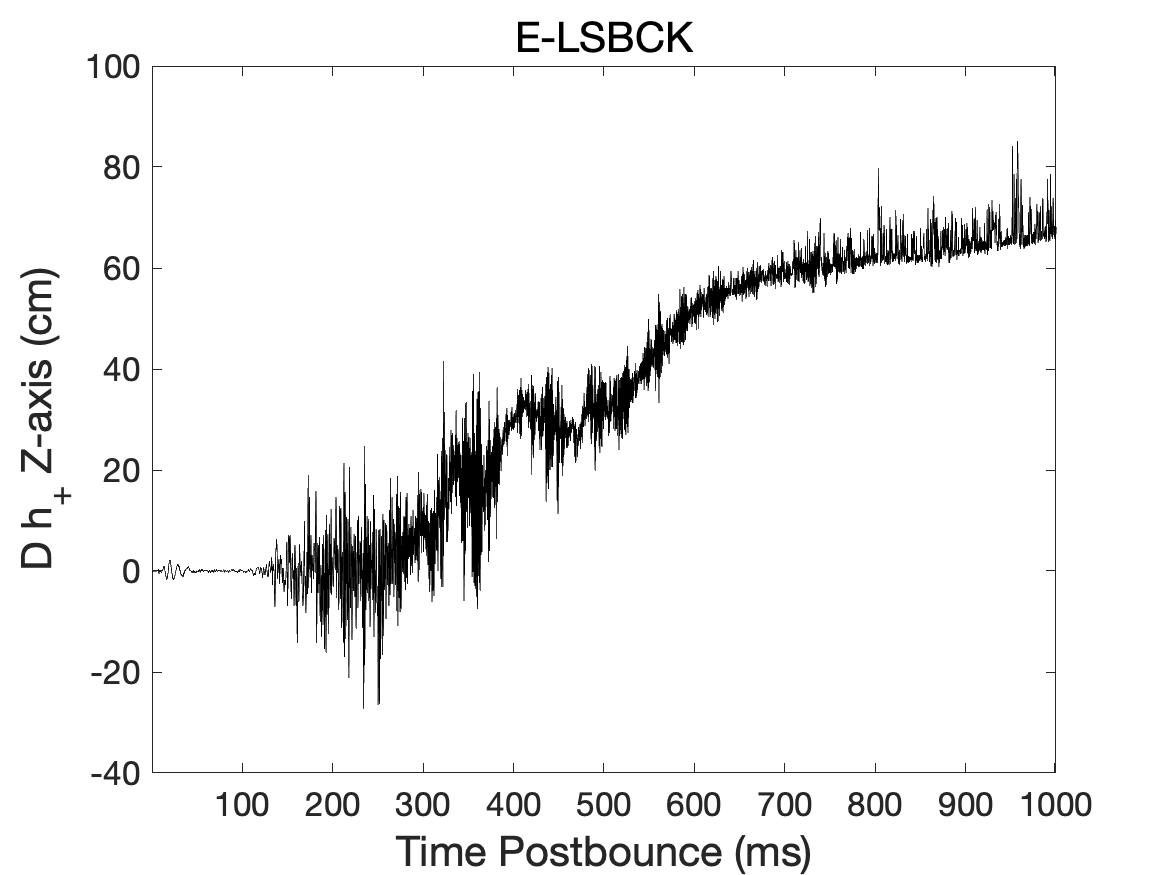}
        \hfill\hfill
        \caption{Strains for each E-series model computed along the z-axis at the source. The small-amplitude, high-frequency signal at late times, superimposed on the low-frequency signals, and the associated noise seen in our spectrograms, are the result of dynamic spatial re-gridding and of no physical consequence. The noise is smallest in E-SFHo and largest in E-LSBCK.}
        \label{fig:strain_all}
    \end{figure*}

\begin{figure*}[!ht]
        \centering
            \includegraphics[width=8.6cm]{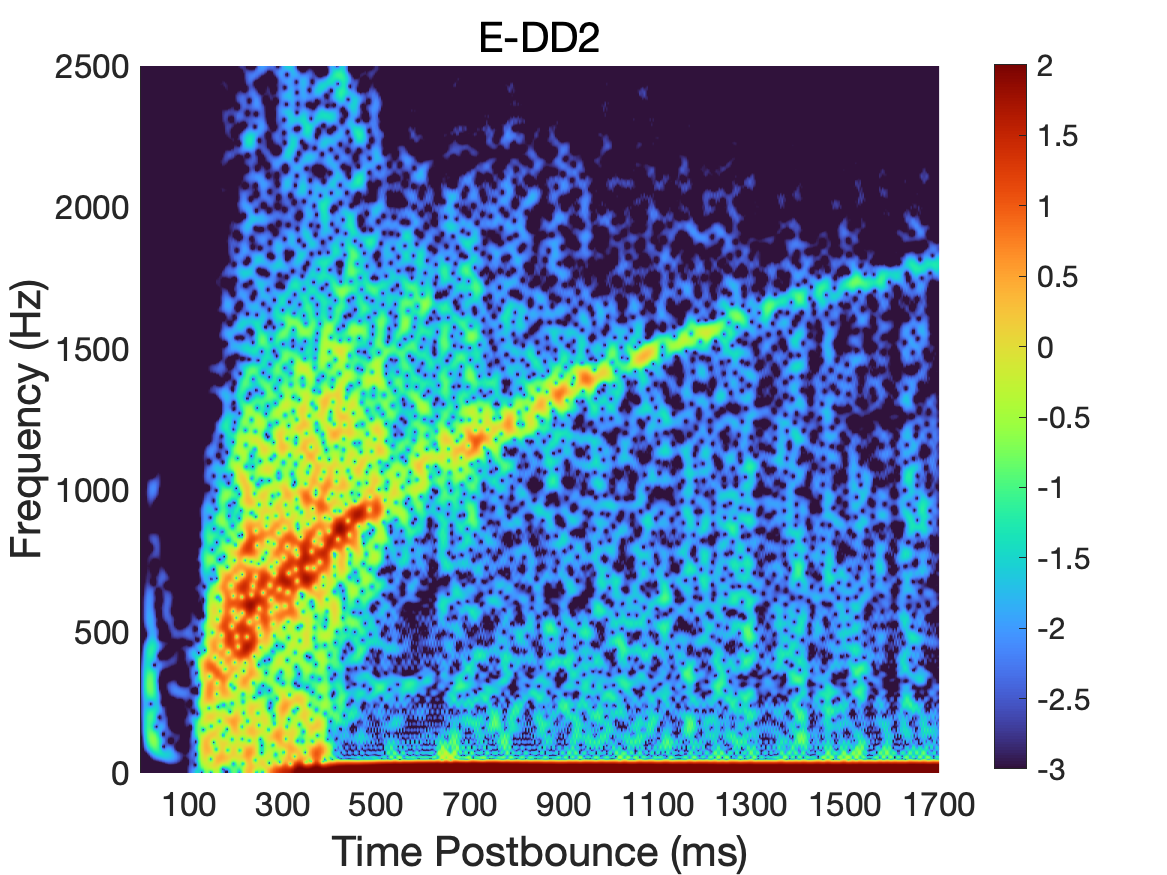}
            \includegraphics[width=8.6cm]{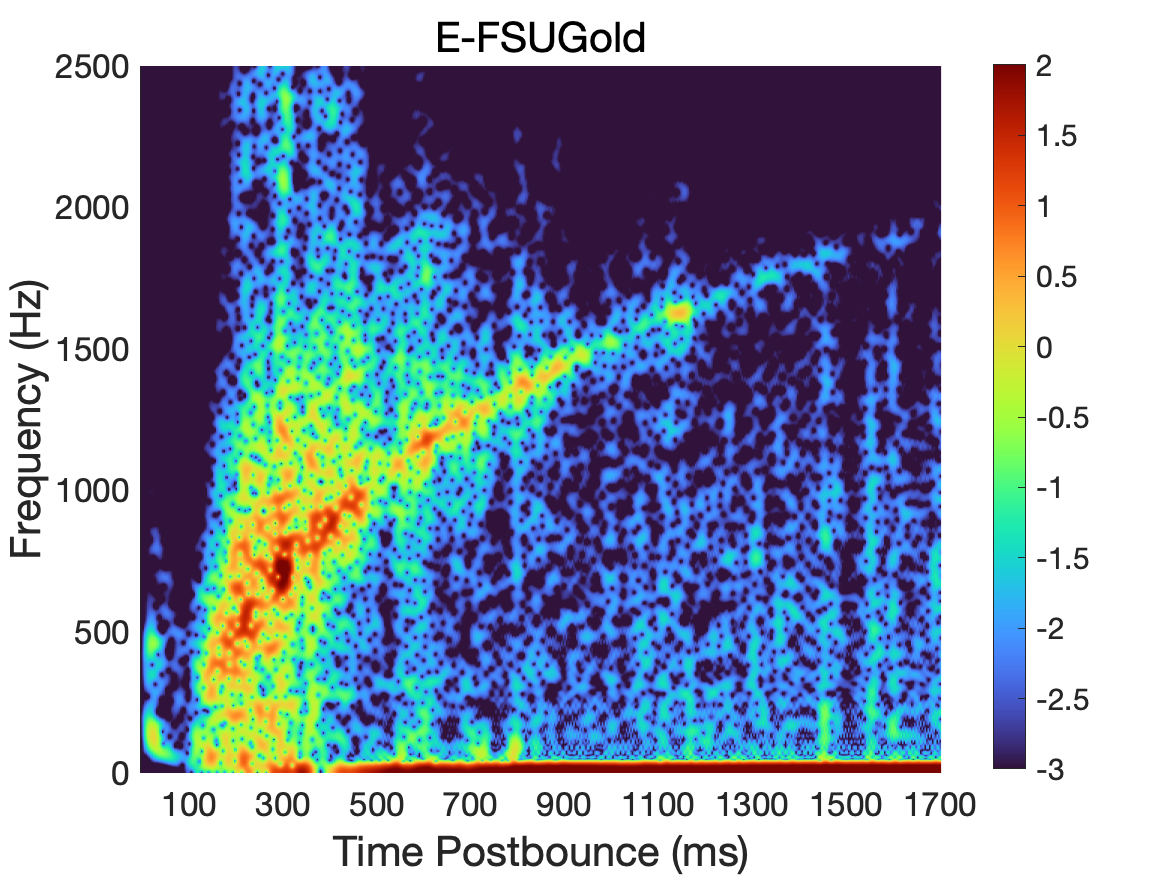}
        \hfill
            \includegraphics[width=8.6cm]{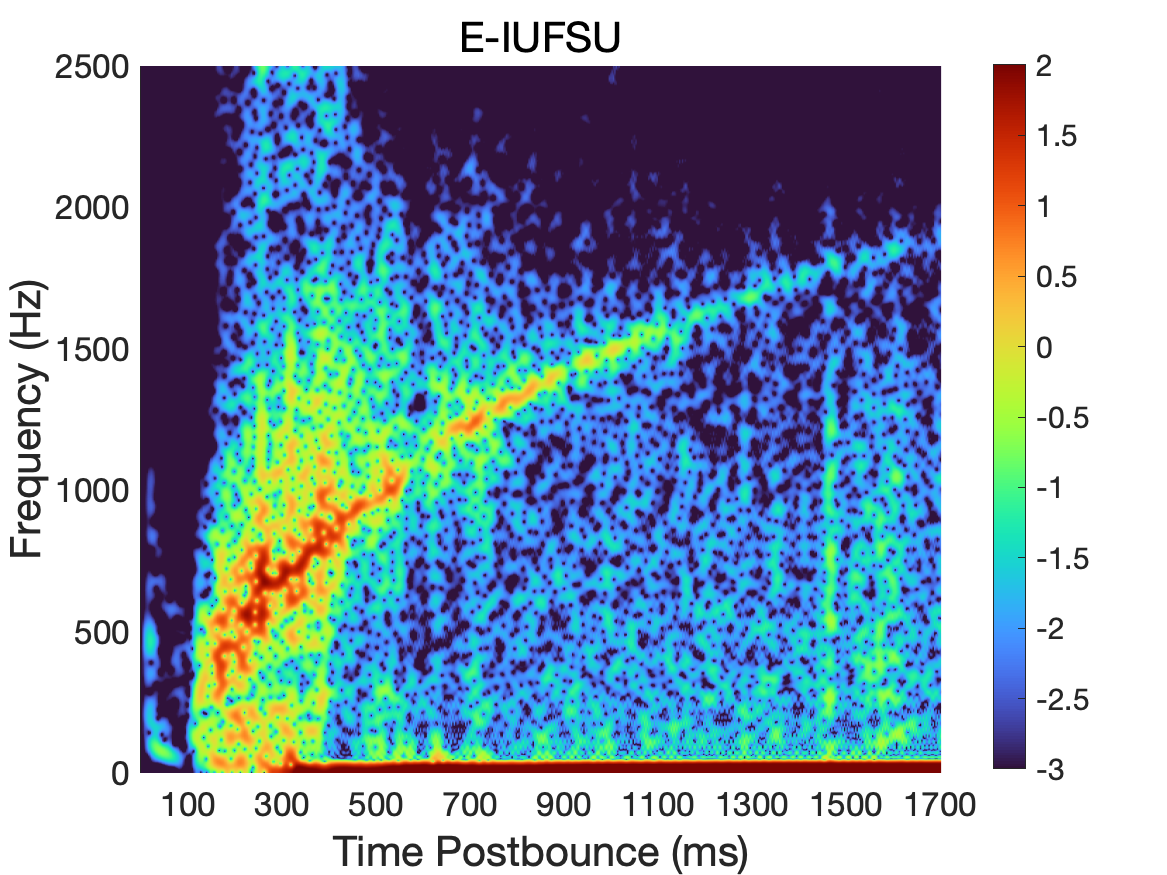}
            \includegraphics[width=8.6cm]{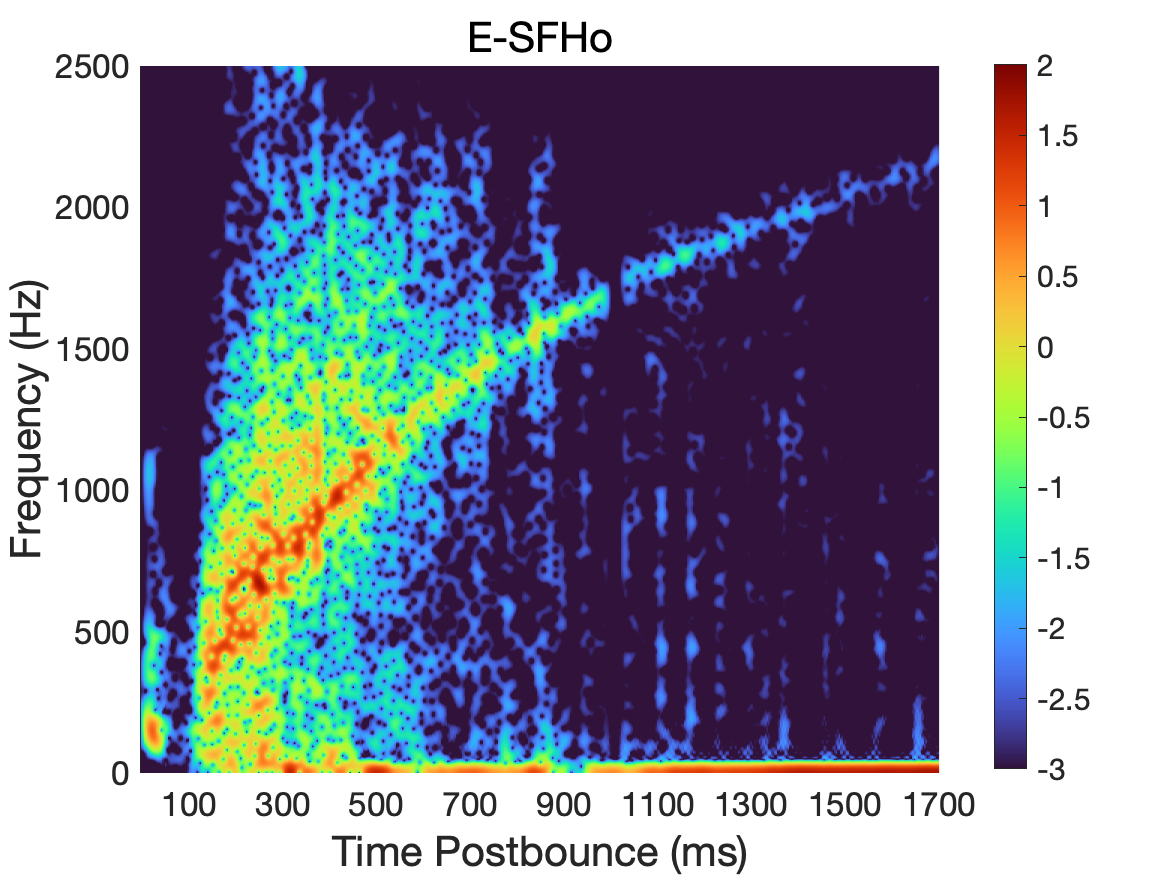}
        \hfill
            \includegraphics[width=8.6cm]{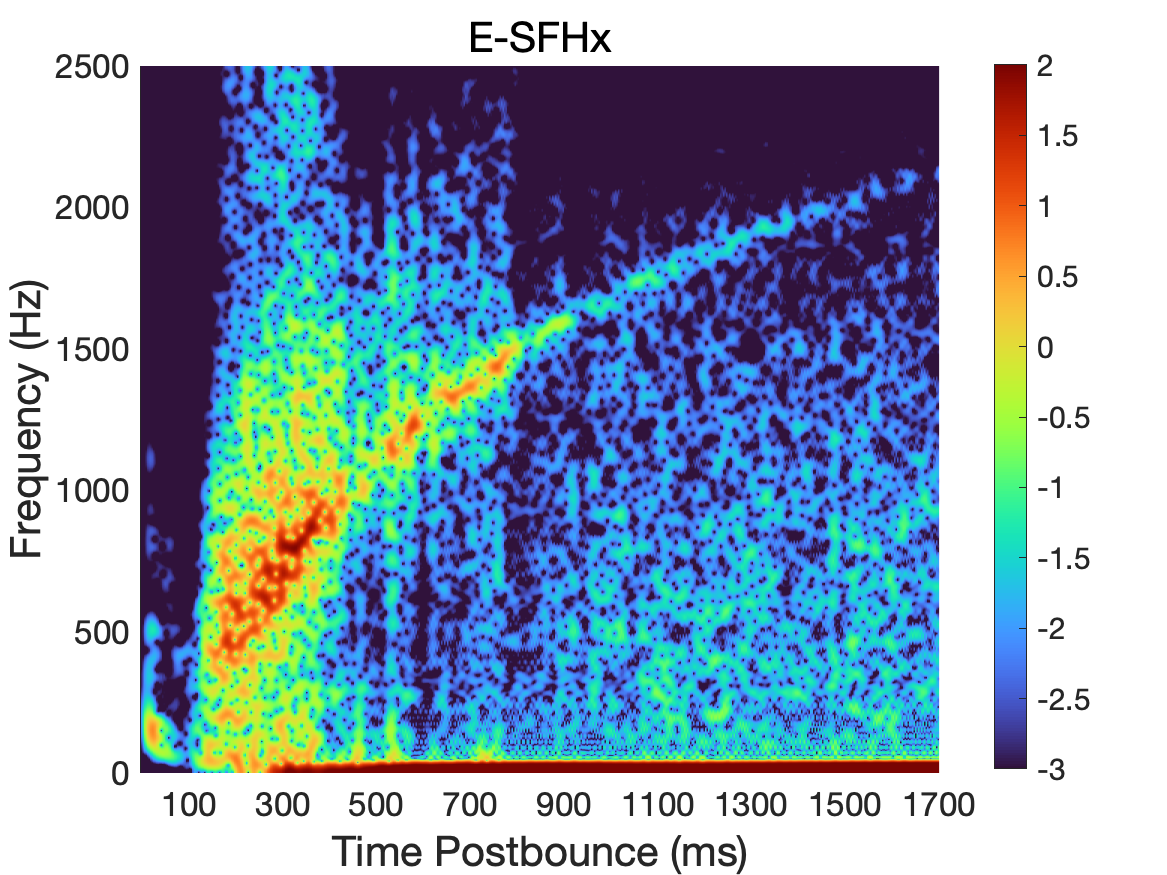}
            \includegraphics[width=8.6cm]{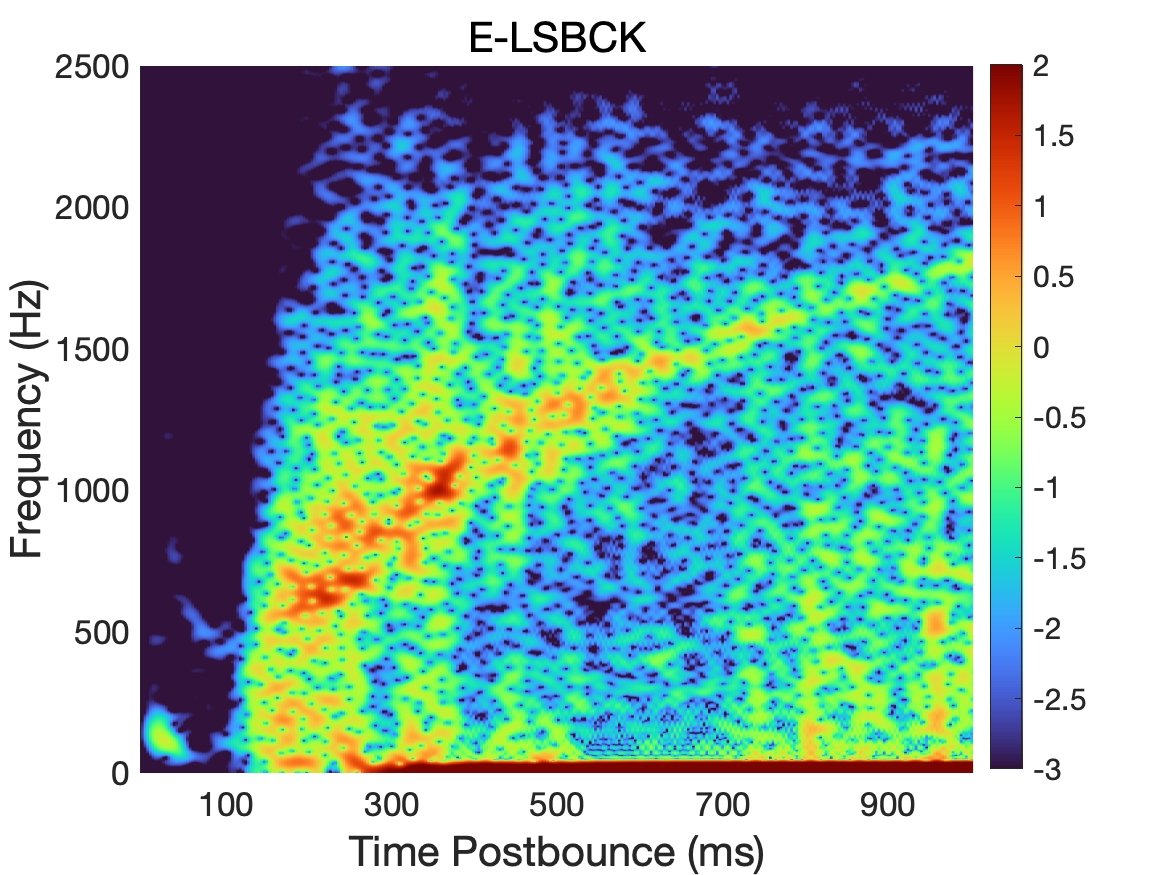}
        \caption{Spectrograms for each E-series model, with the color axis showing the logarithm of the power spectrum, $\mathrm{log_{10}}(P)$.} 
        \label{fig:spec_all}
\end{figure*}

Using the GW extraction method described in Section \ref{sec:GW_Extract} we obtain the GW strains plotted in Figure \ref{fig:strain_all}. Spectrograms of those strains are plotted in Figure \ref{fig:spec_all}, where the color axis shows the logarithm of the power spectrum, $\log_{10}(P)$. We see evidence of all the GW features expected from CCSNe, with the exception of the core-bounce signal due to rotation because these models are non-rotating. We see the effects of prompt convection from 10--60 ms in each simulation. The quiescent period is shown from 60--100 ms by the region where the power spectral density drops below the cutoff value on the colormap, indicating weak GW emission. The HFF starts at some point after 100 ms in the 200--400 Hz range and increases to 1 kHz before 600 ms. Below the HFF, we see the effects of neutrino-driven convection and the SASI, though this is not the focus of this paper. In each model, at the onset of explosion, there is a feature with high power spectral density at frequencies below $\sim50$ Hz, corresponding to  gravitational wave memory developing at the onset of explosion. For each model, the HFF persists into late times in the simulation and has not yet plateaued. This is consistent with a PNS that has not yet reached its final radius. 

\subsection{Fit to HFF}\label{subsec:HFF_lin}
As depicted in Figure \ref{fig:spec_all}, the late-stage behavior of the HFF deviates from linearity. However, Figure \ref{fig:Lmaps} in Section \ref{Sec:HFF_Estimation} will demonstrate that the likelihood of detecting signals above 1 kHz is low given the present sensitivity of the LIGO interferometers.
Hence, our focus, following the approach of \citet{PhysRevD.108.084027}, is on the signal beneath 1 kHz where the HFF shows a roughly linear relationship with time. Additional justification for considering the HFF as linear under 1 kHz is provided in Appendix \ref{app:Curv}.

A key factor in determining the slope of the HFF is the choice of the start time of the HFF. As indicated in \cite{PhysRevD.108.084027}, this choice influences the estimation of the HFF slope, and points to the need to establish a criterion to uniformly define the initial HFF time. Though the stochastic features on the spectrogram are weaker than the HFF, they are not negligible. As a result, the power spectral densities in Figure \ref{fig:spec_all} show signals above the threshold on the color axis in the quiescent period. Including points from this region where the maximum power spectral density rapidly oscillates between frequency bins can strongly impact our estimation of the linear time dependence of the HFF. Therefore, it is crucial to carefully determine the point at which the HFF signal surpasses these signals in strength.

\begin{figure*}[!ht]
        \centering  
            \includegraphics[width=8.6cm]{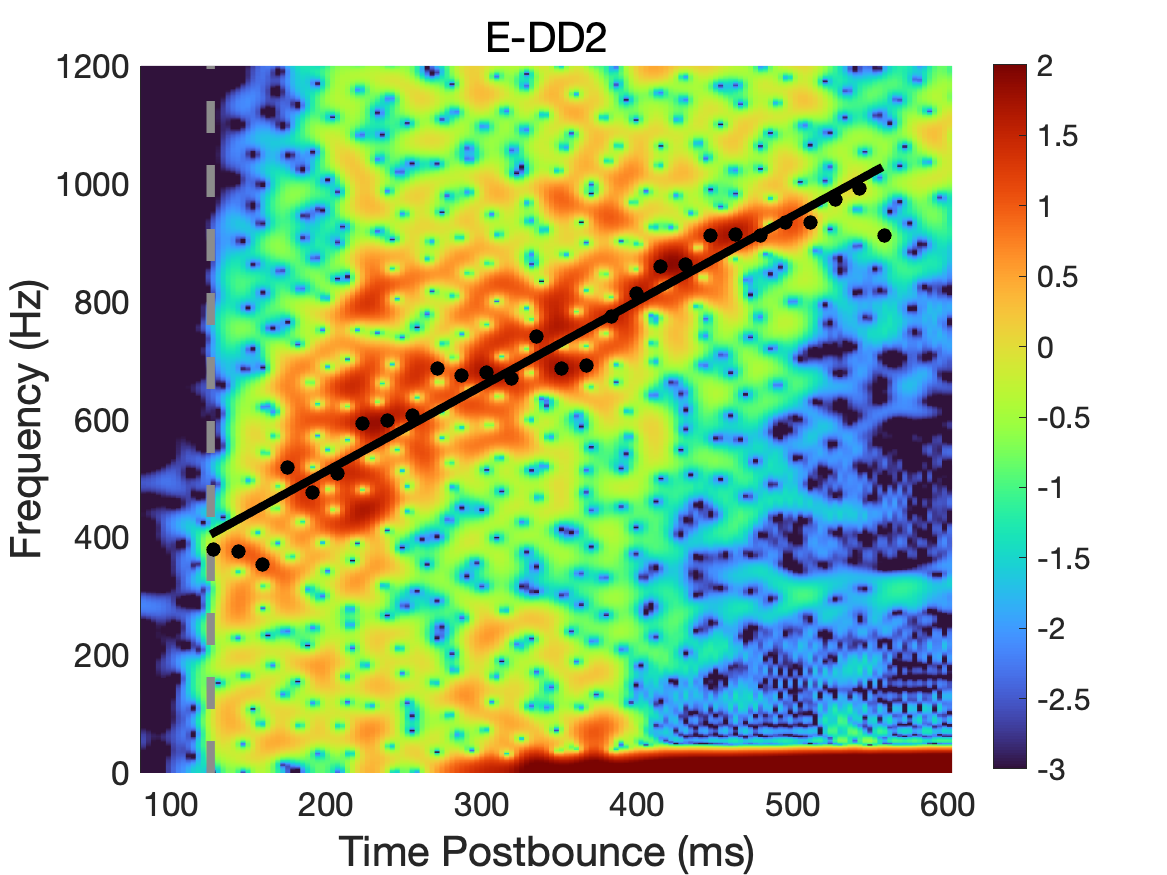}
            \includegraphics[width=8.6cm]{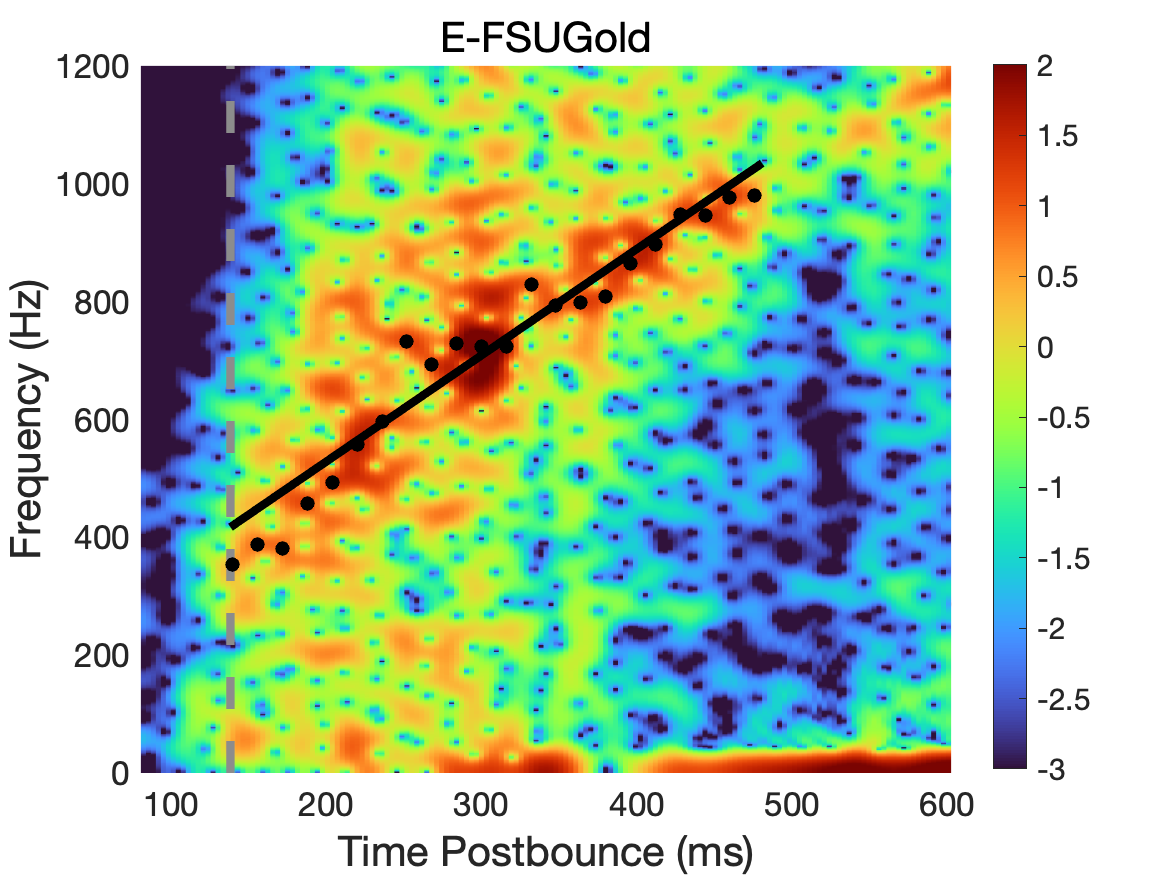}
        \hfill
            \includegraphics[width=8.6cm]{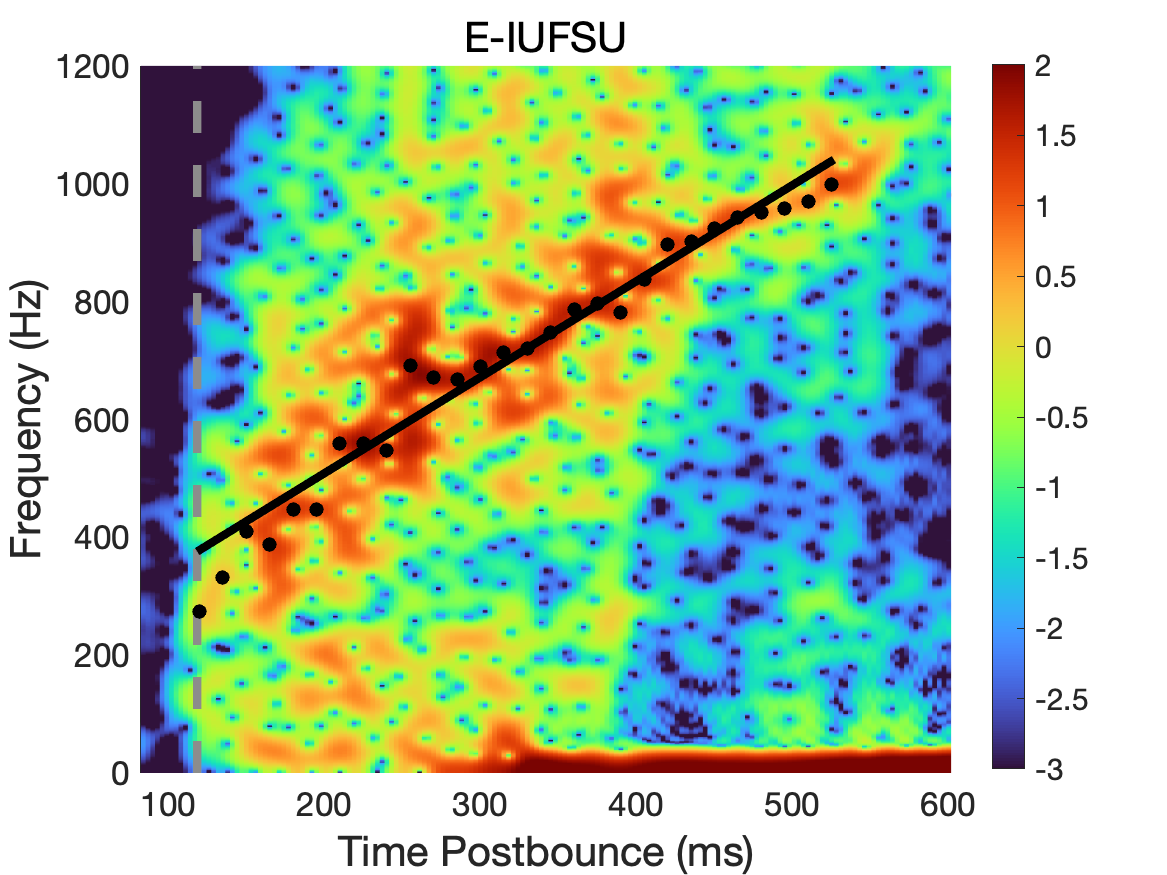}
            \includegraphics[width=8.6cm]{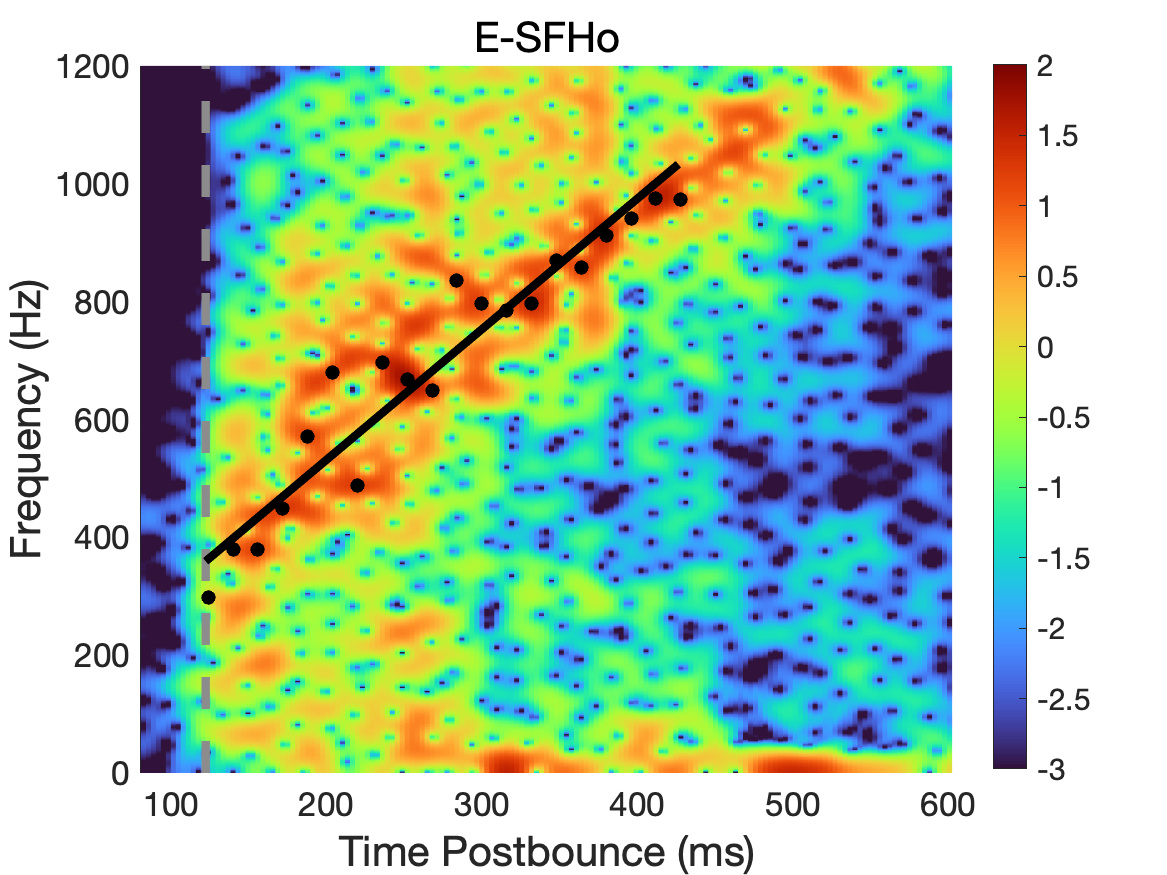}
        \hfill
            \includegraphics[width=8.6cm]{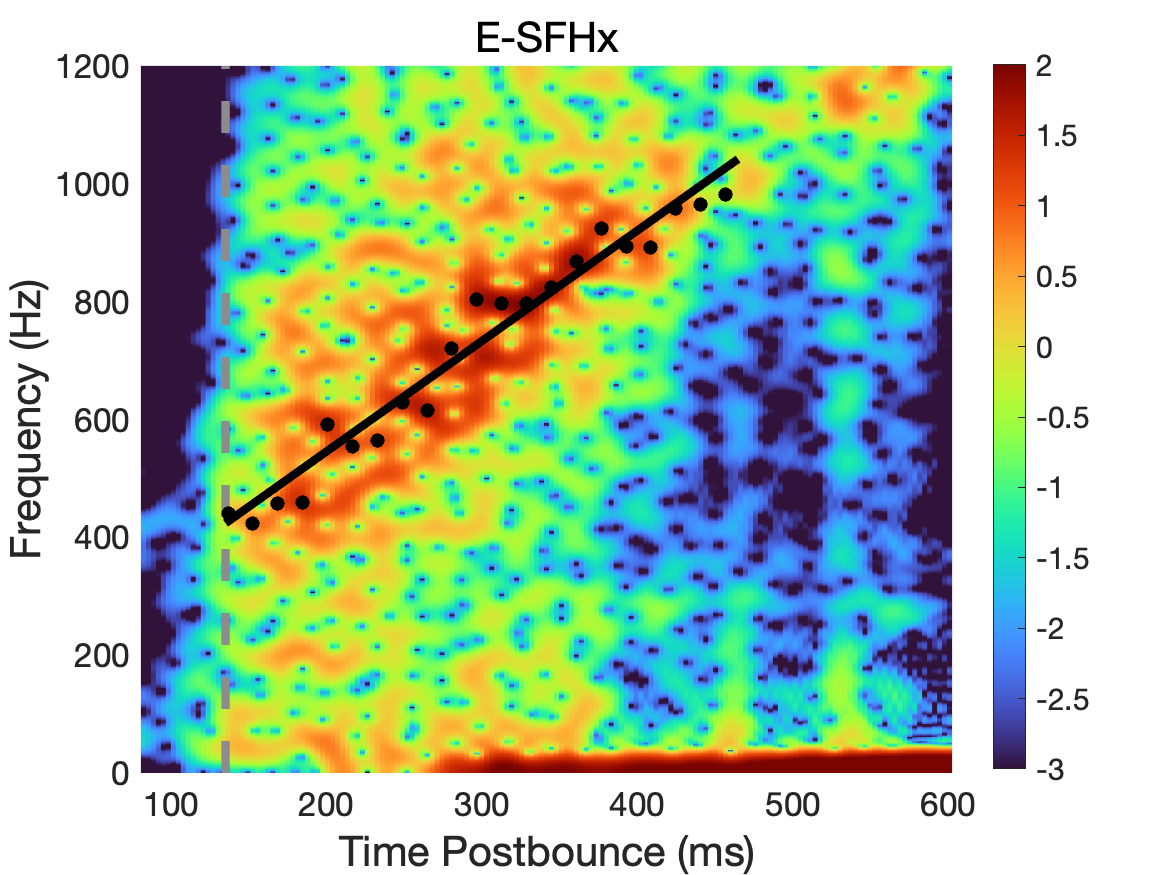}
        \caption{Linear-best-fit lines for maximum power spectral density output. The solid, black line is the first-order approximation, $f=\alpha t +\beta$, of the HFF, as described in the text. The gray, dashed lines represent the starting time of the HFF, \thff.} 
        \label{fig:linear_comp}
\end{figure*}

We define this point as the start of the HFF signal, \thff. If part of the region of the spectrogram in the quiescent period is included in the HFF signal, it can affect the measured slope of the HFF by up to 10\%. In order to compare slope values consistently, we defined \thff\ such that it  avoids this region of the spectrogram without excluding the beginning of the HFF, across all models. To define \thff, we first determine the maximum power spectral density for a single frequency bin above 50 Hz across the entire time domain. This avoids including the high power spectral densities for low frequencies that are related to the GW memory and not related to the HFF. See \citet{Richardson_2024} for a discussion of the contributions to the CCSN GW signal below 50 Hz. We then locate the time at which a single frequency bin above 250 Hz -- \textit{i.e.} the upper limit for SASI based GW signals -- exceeds 1\% of the maximum power spectral density we determined. This frequency bin was chosen by inspection of previous signals to ensure we capture only the HFF signal at early times. Table \ref{tab:comp} shows the starting frequencies of the HFF at \thff, which are all well above the 250 Hz threshold. This provides a clear demarcation between the noisy signal of the quiescent period and the HFF, as is shown by the gray, dashed line depicting \thff\ in Figure \ref{fig:linear_comp}.

\begin{figure*}[!ht]
            \includegraphics[width=8.6cm]{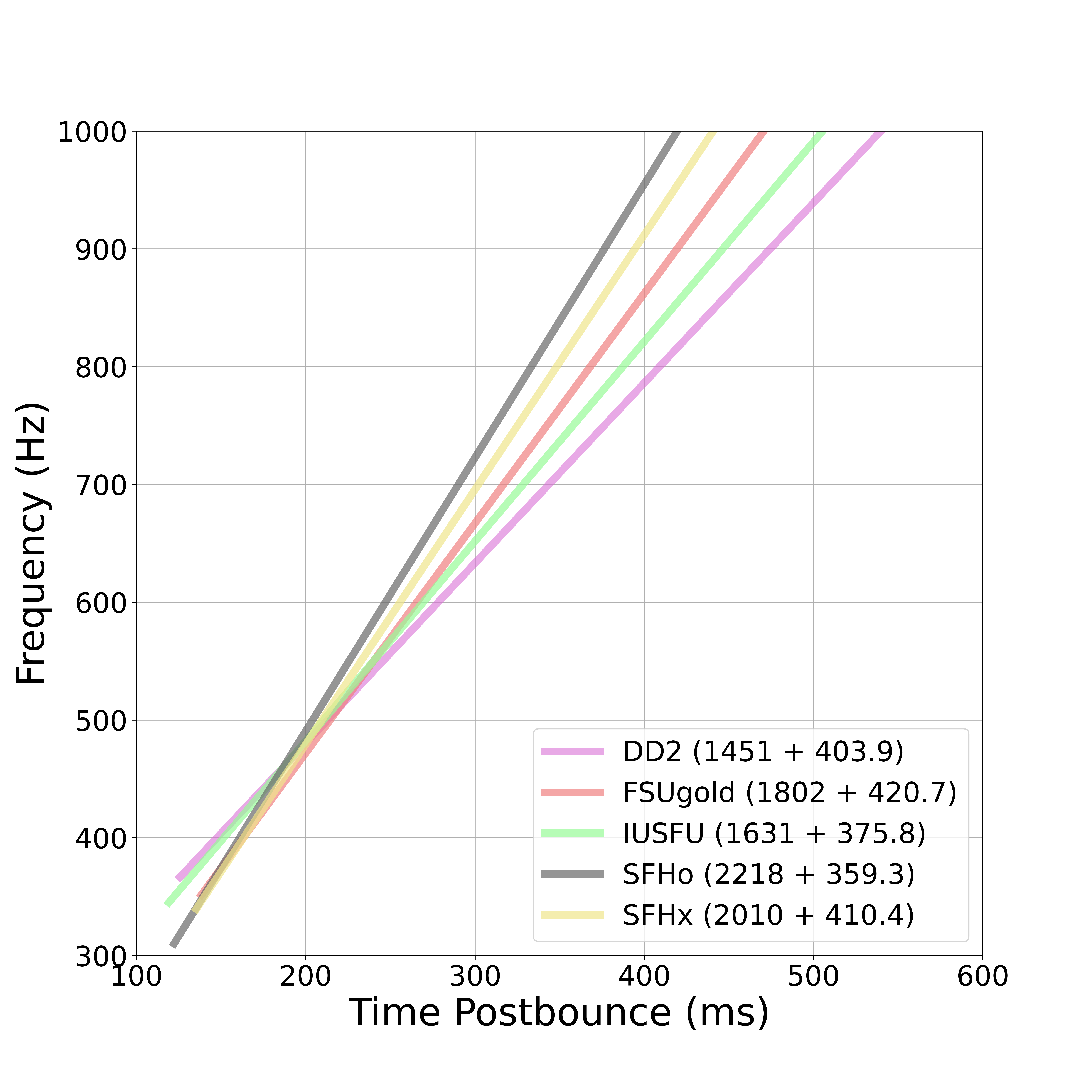}
        \caption{First-order HFF approximations in the absence of noise, across E-series models. } 
        \label{fig:HFF_slopes}
    \end{figure*}

Once the start time of the HFF is determined in the noiseless case, we perform a least-squares regression on the frequency bins of maximum power spectral density for each time bin of the spectrogram, shown in black circles in Figure \ref{fig:linear_comp}. This regression determines an approximation of the HFF of the form $f(t)=\alpha t + \beta$, with $\alpha$ as the linear coefficient and $\beta$ as the starting frequency of the HFF. We note that our regression starts at $t=\thff$ and not $t=0$, which would correspond to the time of bounce. The interpretation of $\beta$ is thus the starting frequency of the HFF, $\beta=f(\thff)$. Figure \ref{fig:HFF_slopes} shows the spectrograms for each model with the linear approximation of the HFF shown as a black line. 

Table \ref{tab:comp} shows the slope of the HFF, starting frequency of the HFF, start time of the HFF (\thff), and maximum absolute percent error (MAPE) of the linear approximation of the observed peak frequencies. There is a clear difference in the slope of the HFF for each of these models. The slope of the HFF could thus be used to further constrain the allowable EOS for the densities and temperatures present during CCSN events. How tightly this coefficient could constrain the allowable EOS will be discussed in section \ref{Sec:Disc}.

    \begin{table*}[!ht]
    \begin{center}
        \begin{tabular}{ l c c c c c c } 
            \hline
            \hline
            EOS & $\alpha$ &  $\beta$ & \thff  & MAPE\\
                &  [Hz s$^{-1}$]  &   [Hz]    &   [ms]    & [\%]\\
            \hline
            DD2 & 1451 & 403.9 & 125.8 & 5.655\\ 
            
            FSUGold & 1802 & 420.7 & 138.4 & 6.198 \\ 
            
            IUFSU & 1631 & 375.8 & 118.9 & 5.906 \\ 
            
            SFHo  & 2218 & 359.3 & 122.3 & 8.135 \\ 
        
            SFHx & 2010 & 410.4 & 135.4 & 4.666\\ 

            LSBCK-2D & 2713 & 394.7 & 128.5 & 7.646\\

            LSBCK-3D & 2727 & 359.3 & 130.1 & 6.074 \\
            \hline
        \end{tabular}
        \caption{HFF first-order polynomial approximation coefficients: $f(t)=\alpha t + \beta$. Second column gives the first-order time dependence of the HFF, \textit{i.e.} linear slope $\alpha$. Third column gives the starting frequency of the HFF at \thff, $\beta$. Fourth column is the start time of the HFF, \thff, as defined in the text. Fifth column represents the maximum absolute percent error (MAPE) of the actual maximum frequencies $A_t$ and the predicted frequencies $F_t$ for $n$ fitted points, $MAPE=100\frac{1}{n}\sum^n_{t=1}|\frac{A_t-F_t}{A_t}|$.}
          \label{tab:comp}
    \end{center}
    \end{table*} 

\subsection{Dimensionality Comparison}
\label{sec:2d3d}

The artificiality of axisymmetry in these two-dimensional models amplifies modes aligned with the symmetry axis, and the inverted turbulent cascade favors the formation of large eddies in post-shock neutrino-driven convection. As a consequence, larger GW strains \cite{Mezz_2020, OcoCou_2018} are produced. Therefore, it is necessary to demonstrate the impact of axisymmetry on the HFF and its detection. Thus, we also conducted an analysis for a single EOS used in \chimera\ simulations in both two and three dimensions. While the three-dimensional simulation was not carried out to late times, the HFF exceeds the 1 kHz threshold within the first few hundred milliseconds.

\begin{figure*}[!ht]
        \centering
            \includegraphics[width=8.6cm]{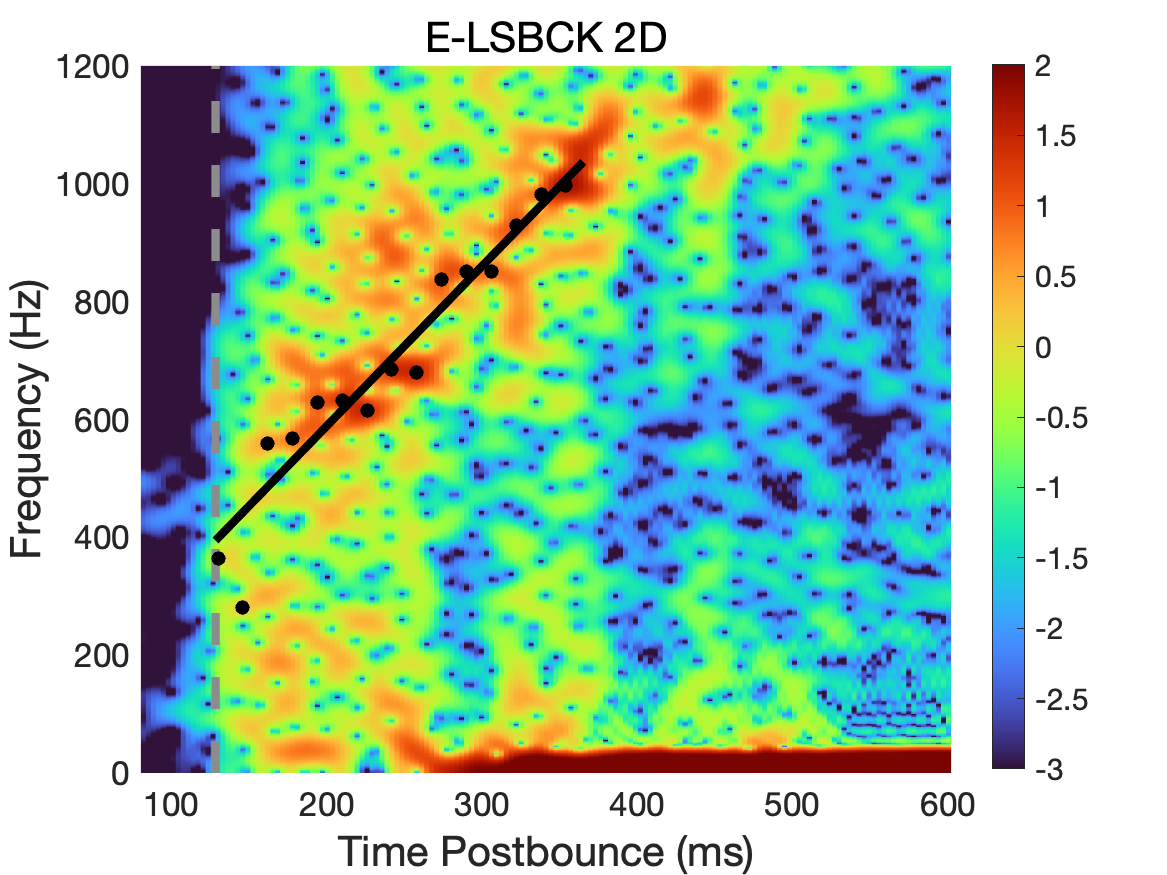}
        \hfill
            \includegraphics[width=8.6cm]{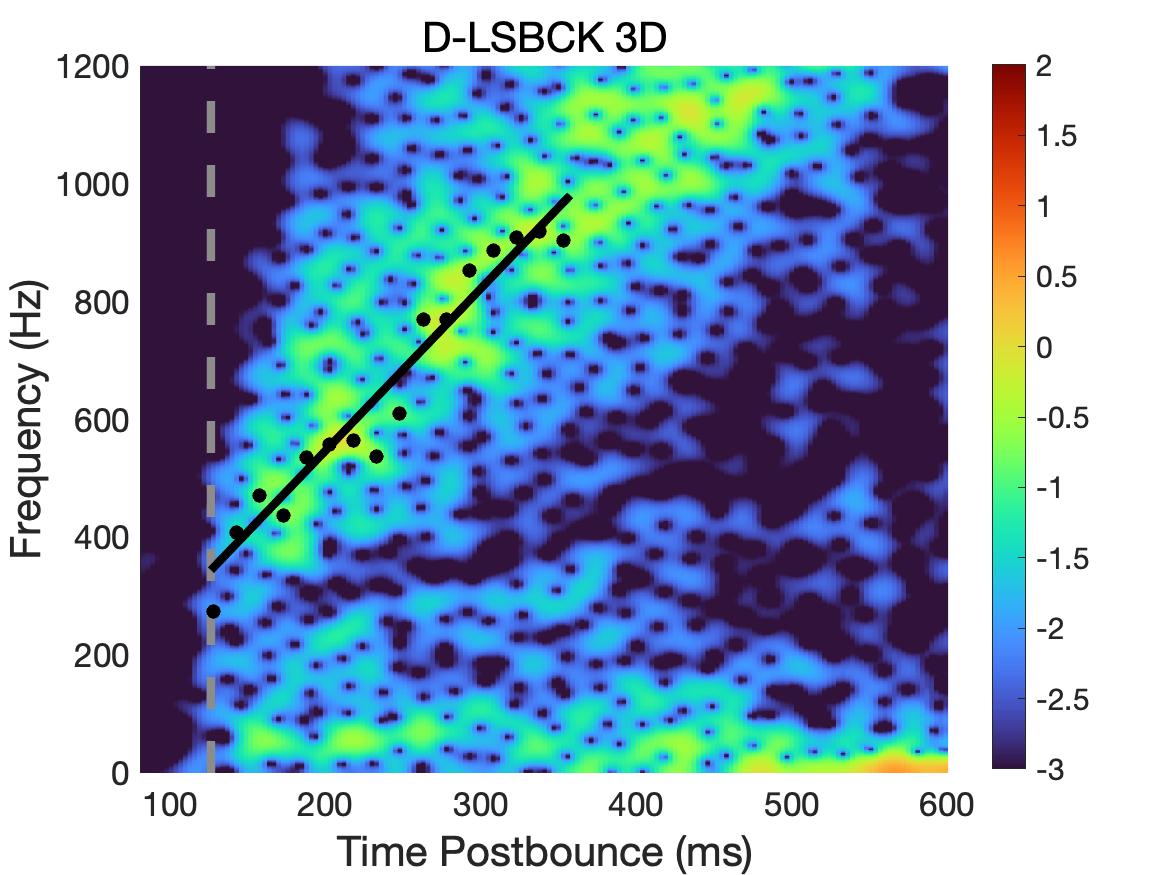}
  \caption{Spectrograms for two-dimensional (left) and three-dimensional (right) CCSN simulations using the LSBCK EOS. Spectrograms were produced using the same parameters used in Figure \ref{fig:linear_comp}, the black line shows the first-order approximation of the HFF, and the gray, dashed line represents \thff\, as defined previously.}
\label{fig:LSBCK_2D_3D}
\end{figure*}

In Figure \ref{fig:LSBCK_2D_3D}, the spectrograms for the two- and three-dimensional simulations using the LSBCK EOS are shown. The beginning of the HFF was defined with \thff\, and the HFF polynomial fit was computed as described in Section \ref{subsec:HFF_lin}. The two-dimensional spectrogram shows an HFF polynomial approximation with a slope of 2713~Hz~s$^{-1}$, starting HFF frequency of $395$ Hz, and \thff\ of $0.1285$ s. The three-dimensional spectrogram shows an HFF polynomial approximation with a slope of 2727 Hz~s$^{-1}$, starting HFF frequency of $359$ Hz, and \thff\ of $0.1301$ s. These results are promising, showing that between two- and three-dimensional simulations the effect on the measurable properties of the HFF seem minimal. It also shows the consistency of the definition of \thff, differing only by $1.6$ ms. However, the color axis on each spectrogram does show significant differences in the strength of the signal. The mean power spectral density across the spectrogram for the two-dimensional GW signal is $\sim 8.4$ times larger than the mean power spectral density across the spectrogram for the three-dimensional GW signal. This means that the properties of the HFF should be consistent between two- and three-dimensional simulations in the noiseless case, but that real CCSN events may be weaker than the signals that give the results we present in the following section. 

\section{HFF Detection and Analysis with Noise}\label{Sec:HFF_Estimation}

Following the approach developed in \cite{PhysRevD.108.084027} to predict the slope of the HFF
we chose a Deep Neural Network (DNN) for our study because it is specifically designed to learn from both linear and non-linear relationships present in the input data. Additionally, it has the ability to automatically extract features from the input domain, resulting in high performance (see for example \cite{2021PhRvD.103f3011L}). Compared to other regression methods in the field of data analysis, the DNN offers this high performance at a relatively low computational cost. In order to assess the accuracy of the DNN algorithm, we utilized various performance metrics, which are described later in this section.
Our approach for analyzing the CCSN GW signals from the E-series with interferometric noise is described as follows:
(i) cWB event production: we use the cWB algorithm to create likelihood time-frequency maps, $L$, for the events that have been identified through an event production analysis. The cWB configuration parameters mirror those employed in \cite{PhysRevD.108.084027}, ensuring consistency in the operation of the pipeline.
(ii) Likelihood Map Processing: we process the likelihood time-frequency maps to extract the physical information embedded in the detected CCSN GW events (see Section II-C in \cite{PhysRevD.108.084027}).
(iii) DNN Model: our primary tool for estimating the HFF slope is the DNN as described in \cite{PhysRevD.108.084027}.
For a comparison of different attempts to perform parameter estimation for core-collapse supernovae we refer the reader to \cite{2022PhRvD.105f3018P, 2018PhRvD..98l2002A}.

\subsection{cWB Event Production Analysis}\label{subsec:cWB}

The cWB algorithm operates through a three-step process: first, it scans for coincident signal power by utilizing the Wilson-Daubechiers-Meyer transform to project multi-detector data onto the wavelet domain~\cite{Necula_2012}; second, it identifies coherent time-frequency components with amplitudes surpassing noise levels; and finally, it clusters these components to generate a likelihood time-frequency map denoted $L = {(t_i, f_i), l_i}_{i=1}^{N_L}$. Here, $l_i$ represents the value of the likelihood point in time ($t_i$) and frequency ($f_i$), with $N_L$ indicating the total number of time-frequency points. This map provides the essential time-frequency data that are used to reconstruct the detected GW signal and is used to estimate the slope of the HFF present in CCSN GW signals. 

\begin{figure*}[!ht]
        \centering
            \includegraphics[width=5.6cm]{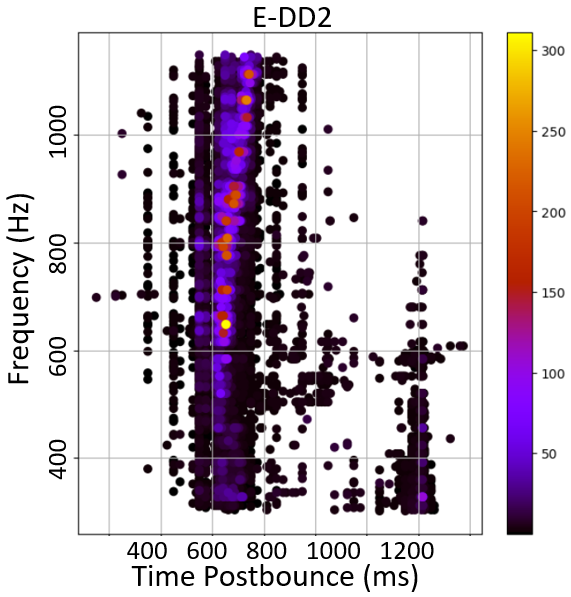}
        \hfill
            \includegraphics[width=5.6cm]{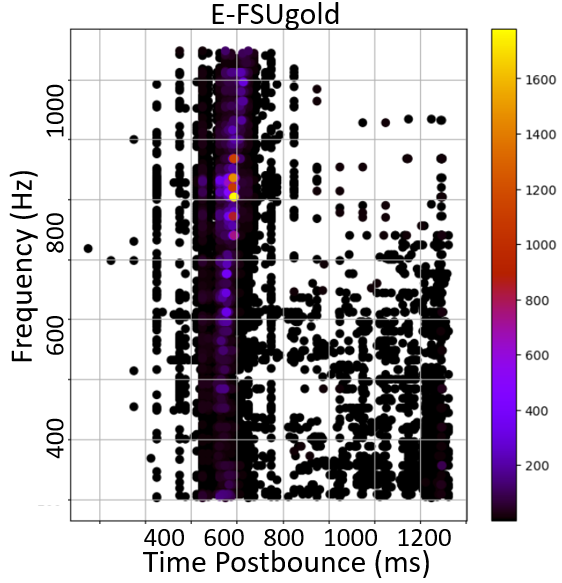}
        \hfill
            \includegraphics[width=5.6cm]{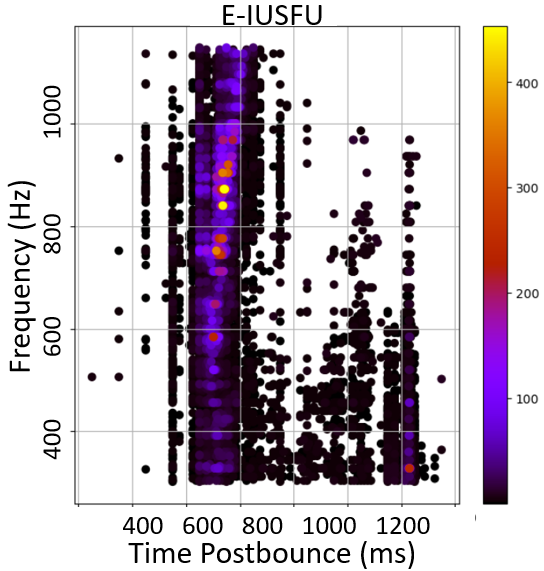}
        \hfill
            \includegraphics[width=5.6cm]{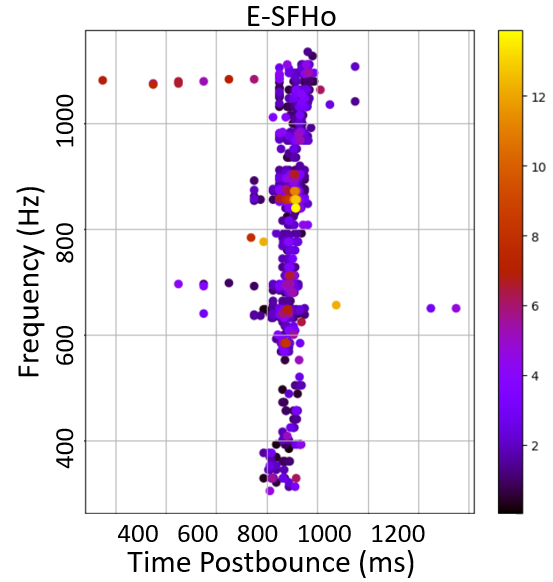}
        \hfill
            \includegraphics[width=5.6cm]{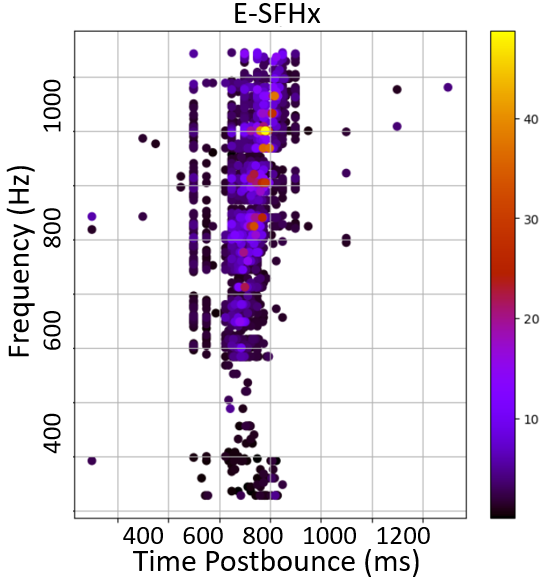}
        \caption{Examples of time-frequency likelihood maps $L$ at 1 kpc associated with each E-series model. The color map represents the likelihood of a point being considered a detection, with black being the lowest likelihood considered above the noise level and yellow being the point with the highest likelihood of detection. } 
        \label{fig:Lmaps}
    \end{figure*}

For our analysis, the cWB event production was set up with a standard configuration during the second half of the third observing run (O3b) \cite{KAGRA:2023pio}. Then, the CCSN GW signal was injected into this detector strain data -- \textit{i.e.} real interferometric noise -- every 50 seconds at distances of 1, 5, and 10 kpc. Each signal was treated as traveling in an equatorial orientation with the detector. In Figure \ref{fig:Lmaps}, we show examples of likelihood time-frequency maps for the E-series GW signals detected in real interferometric noise at a distance of 1 kpc.

The number of detections changed depending on the Galactic distances and cWB operation parameters; we obtained 1513, 1421, and 1383 detections for the E-series GW signals at 1 kpc, 5 kpc, and 10 kpc respectively. The configuration file, which controls the operation of cWB for event production, included the optimal probability for black pixel selection ($bpp$) set at 0.05, the subnetwork threshold ($subnet$) set at 0.5, the production thresholds ($netRHO$ and $netCC$) set at 4.0 and 0.4, respectively, and the lowest and highest frequencies adjusted to 250~Hz and 1000~Hz, respectively.

\subsection{Likelihood Maps Processing}\label{subsec:L}

The likelihood time-frequency maps, $L = { (t_i,f_i) ,l_i }_{i=1}^{N_L}$, exhibit variations in the number of points, frequency range, and time interval between detected events. To facilitate the input for the machine learning model tasked with estimating the slope of the HFF, it is imperative to create a standardized data representation. To do this, we define a function denoted $f(\cdot)$ that transforms $L$ into a two-dimensional data matrix $X(t,f)$, where both the width ($t$ dimension) and the height ($f$ dimension) remain constant for any identified GW event.

Given $L = { (t_i,f_i) ,l_i }{i=1}^{N_L}$, the image construction involves the following procedure. Initially, the time-frequency point with the maximum likelihood value, denoted $\{ t_{m}, f_{m} \}$, is selected. Subsequently, a region around $t_{m}$ is chosen within the interval $[t_{m}-\delta_t, \ t_{m}+\delta_t]$. At the same time, a region around $f_{m}$ is selected within the interval [250~Hz, 1000~Hz], where the reconstructed signal from the coherent waveBurst (cWB) is more accurate and approximately linear. Here, the length of $\delta_t$ is fixed at 0.3~s to ensure that the time interval is large enough to cover the primary evolution of the HFF present in the CCSN GW signal considered in this study.

\subsection{A Deep Neural Network Model for HFF Slope Estimation in Real Interferometric Data}\label{subsec:DNN}

The DNN for regression modeling, implemented to estimate the slope of the HFF, utilizes rectified linear unit (ReLu) activation functions in the hidden layers and a linear function in the output layer, which allows for the regression stage. Training of the model involved fitting the synaptic weights using the back-propagation learning algorithm with the root-mean-squared propagator (MNSprop) as the loss function, a learning rate of 0.001, a batch size of 512 samples of the training data, and 300 epochs. The DNN model encompasses 64, 32, and 16 neurons in the hidden layers, resulting in 64,773 weights being adjusted during the learning process.

\begin{figure*}[!ht]
        \centering
            \includegraphics[width=8.6cm]{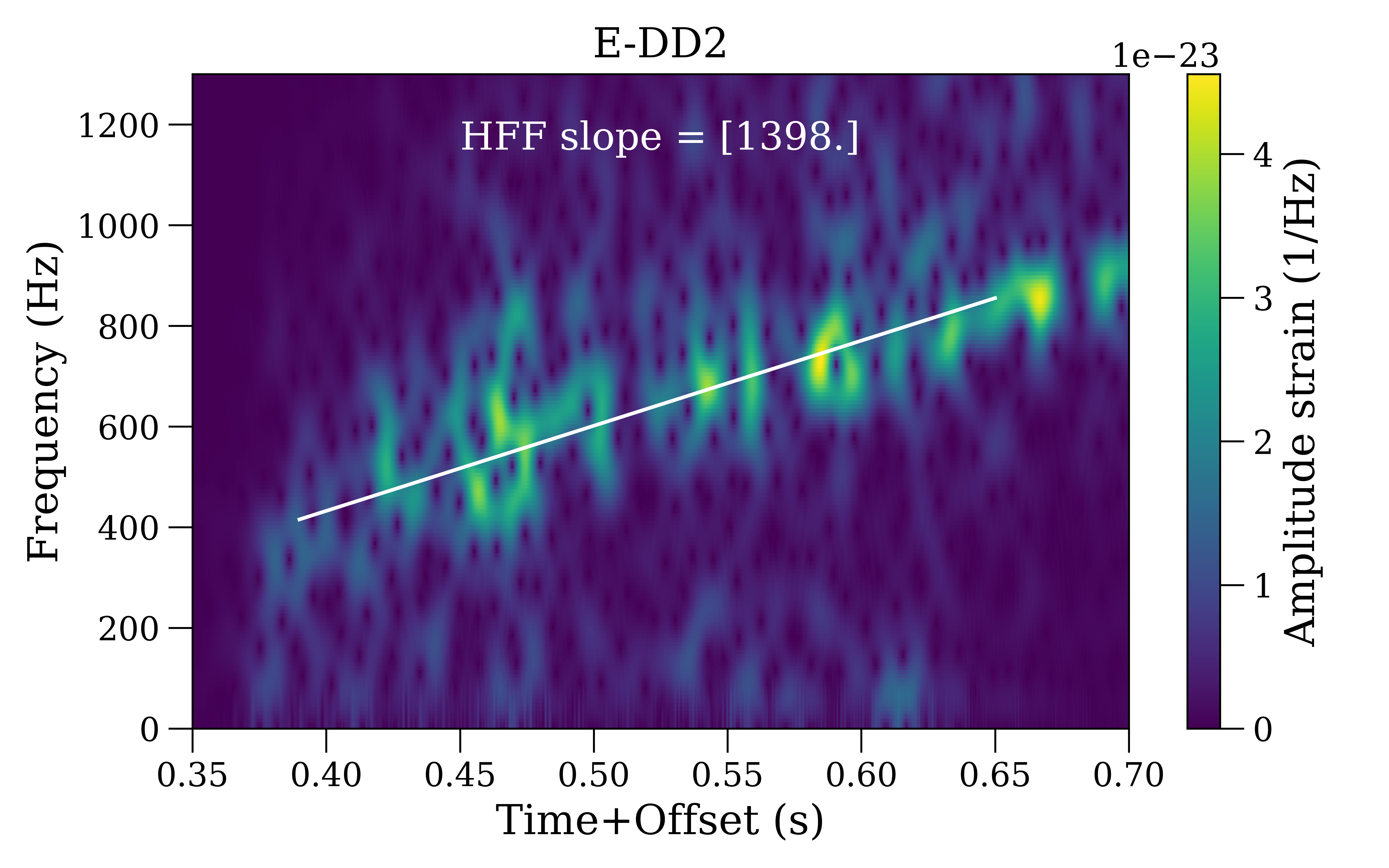}
        \hfill
            \includegraphics[width=8.6cm]{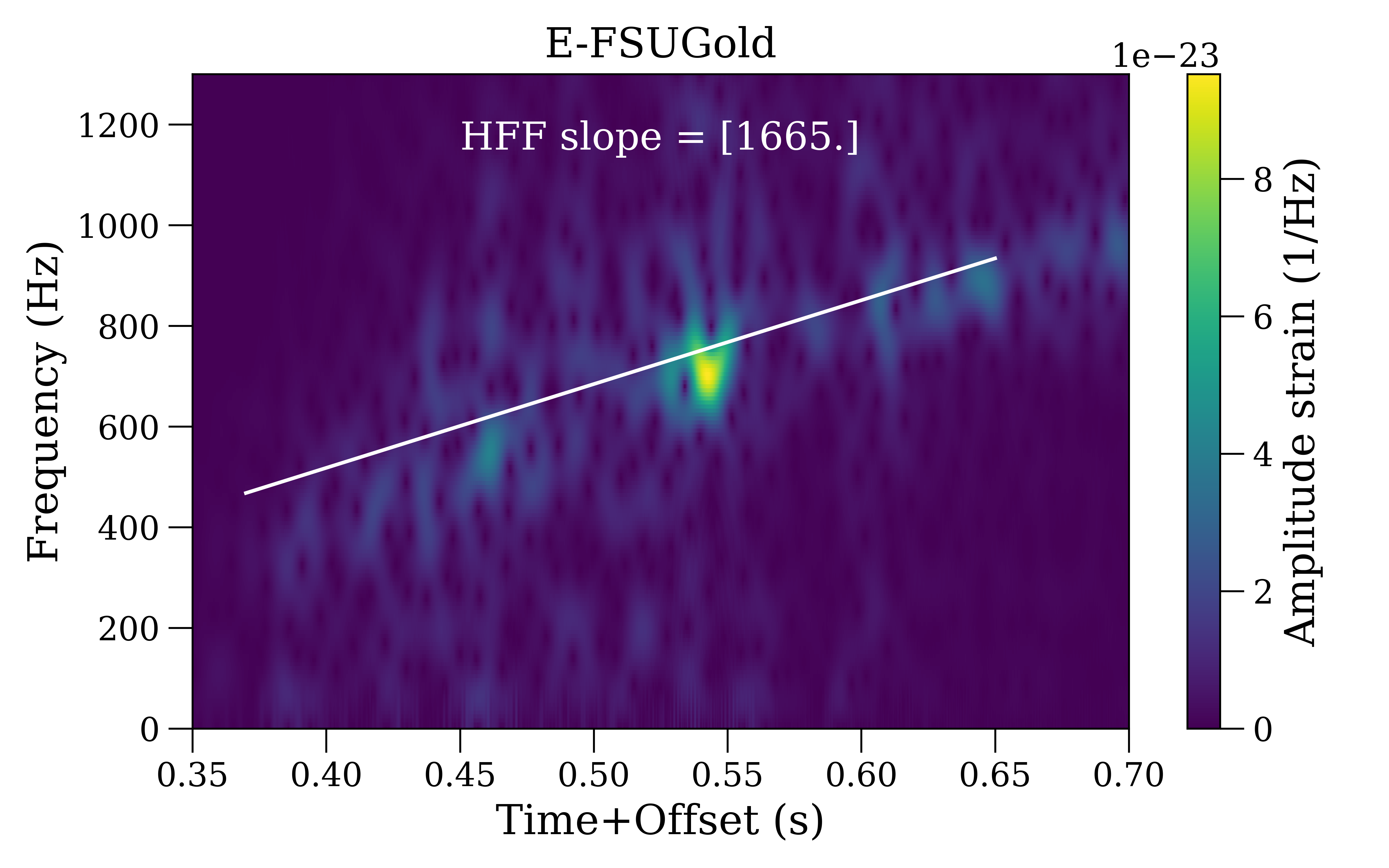}
        \hfill
            \includegraphics[width=8.6cm]{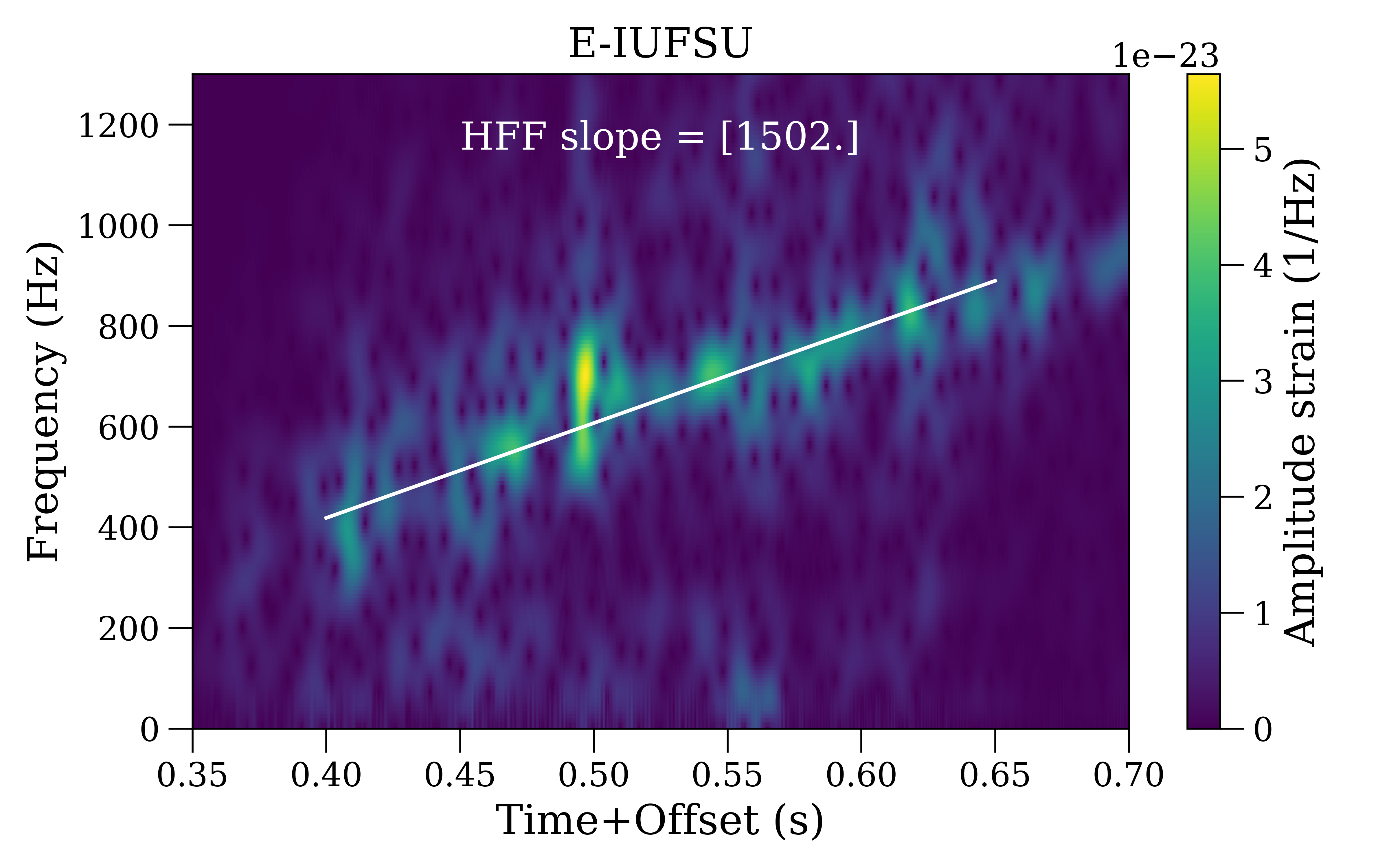}
        \hfill
            \includegraphics[width=8.6cm]{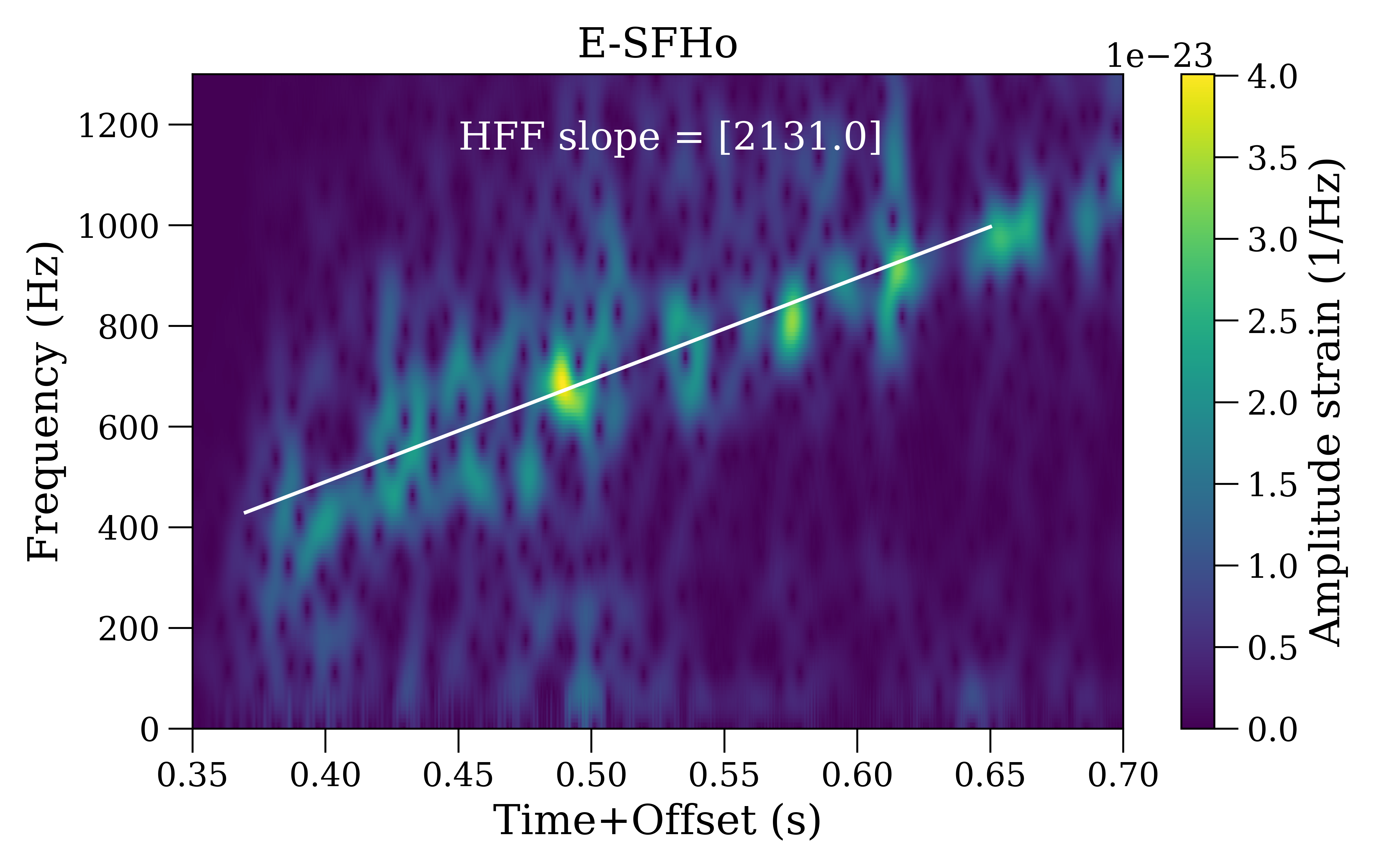}
        \hfill
            \includegraphics[width=8.6cm]{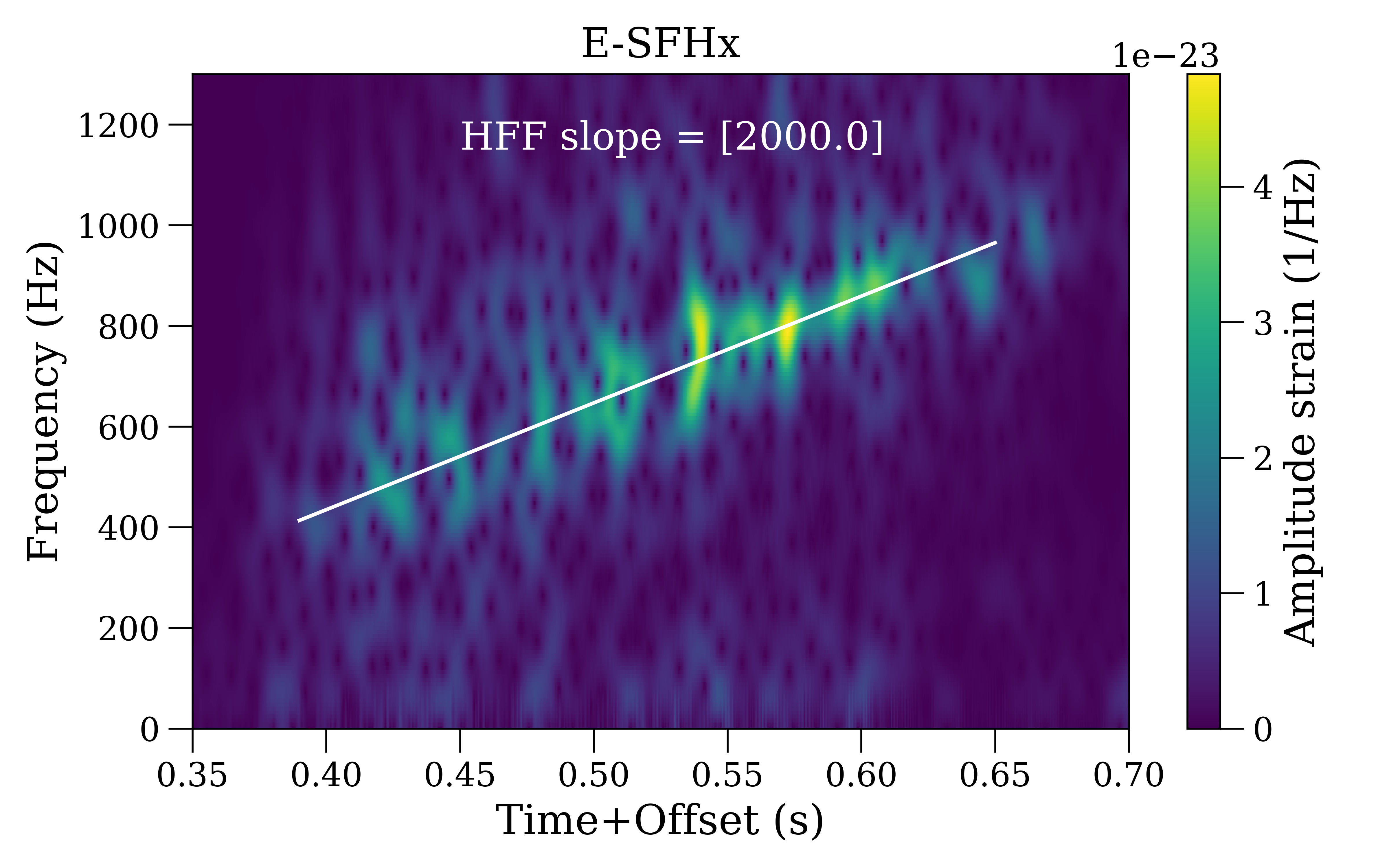}
        \caption{Spectrograms for each E-series model in the absence of noise, with the signal up-sampled to standard LIGO detection frequency of 16,384 Hz. The solid, white lines in the spectrograms trace the increasing frequency over time and its corresponding HFF slope estimation using the methodology introduced in \cite{PhysRevD.108.084027}. These slopes are around 8.5\% higher than those shown in Figure \ref{fig:HFF_slopes} for the E-FSUGold and E-IUFSU models. The sensitivity of the slope of the HFF to \thff\ and differences in sampling rate causing slight variations in \thff\ are the reasons for this.} 
        \label{fig:HFF_slopes_base}
    \end{figure*}

To train the DNN model, we constructed a training data set, called $\mathcal{D}_{\rm train}$, which comprises synthetic CCSN GW signals (see \cite{PhysRevD.108.084027} Section II-A). To test the algorithm, we used the E-series GW signals. This serves as our test dataset, denoted $\mathcal{D}_{\rm test}$, and includes estimated values of the slope of the HFF under noise-free conditions, denoted $s$. The computation of $s$ follows the same procedure as outlined in Section \ref{subsec:HFF_lin}. The CCSN GW signal linked to each EOS was intentionally withheld from the training phase, creating unknown territory for the DNN architecture to estimate the HFF slope. In Figure \ref{fig:HFF_slopes_base}, we show the resulting first-order approximation of the HFF for the E-series in the absence of noise, using spectrograms produced at the LIGO sampling frequency of 16,384 Hz.

\begin{figure*}[!ht]
        \centering
            \includegraphics[width=8cm, height=4cm]{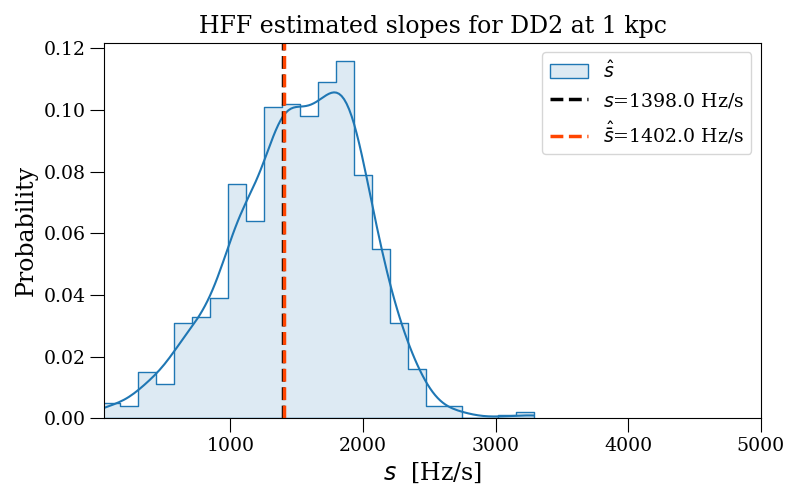}
        \hfill
            \includegraphics[width=8cm, height=4cm]{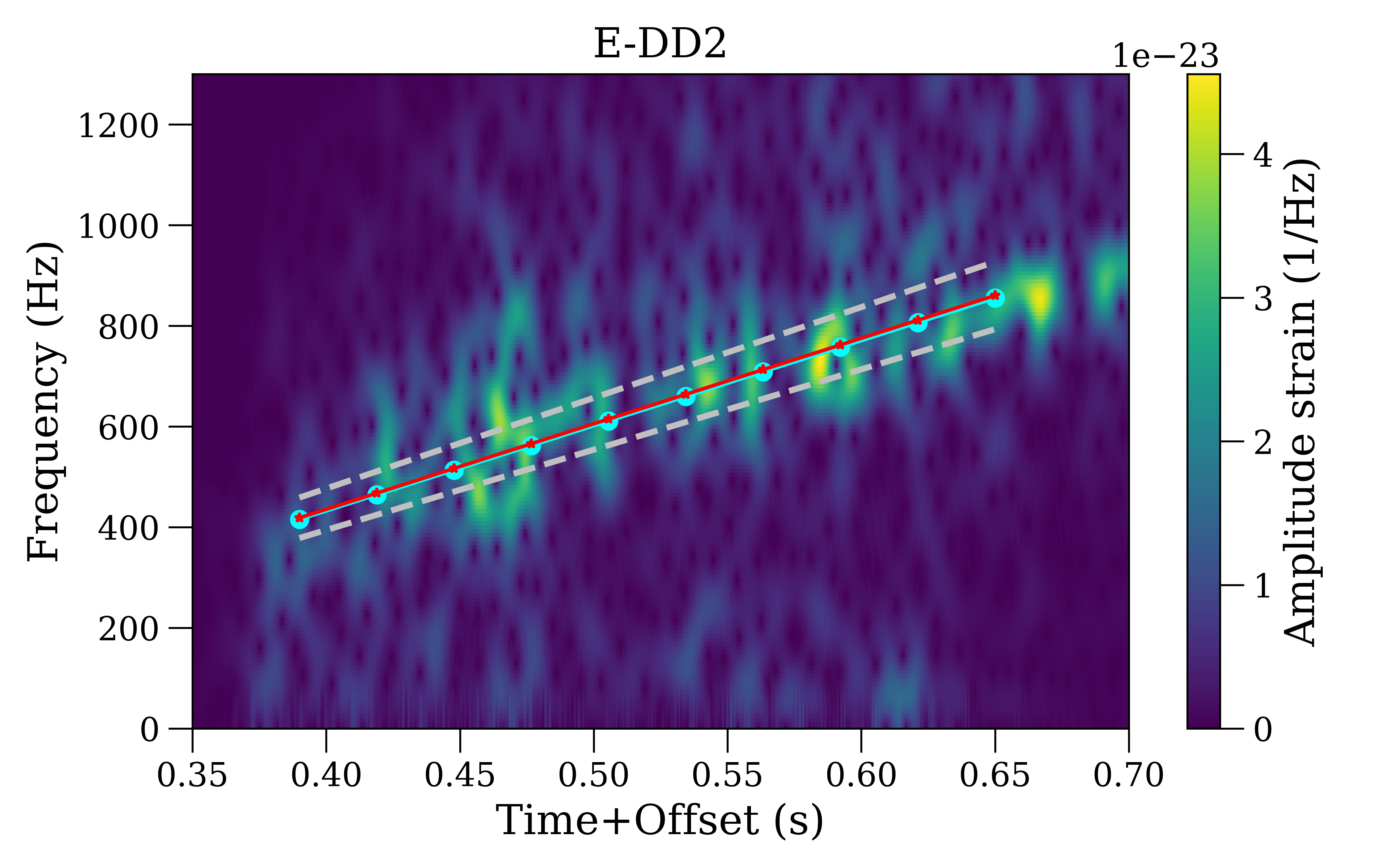}
        \hfill
            \includegraphics[width=8cm, height=4cm]{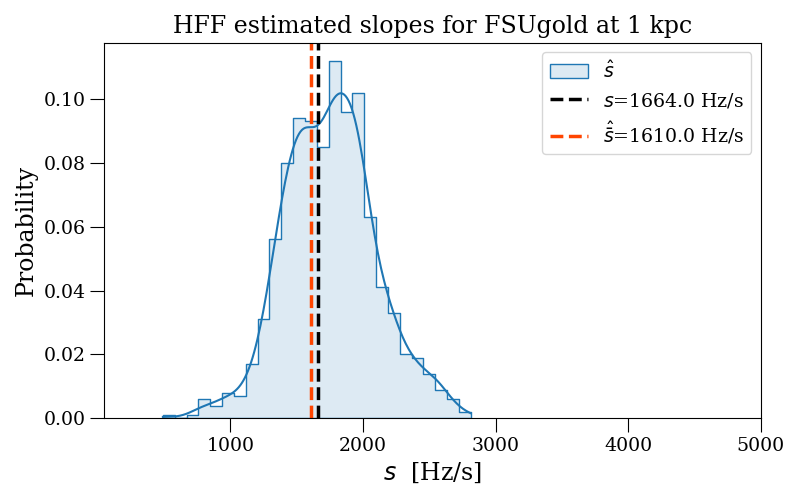}
        \hfill
            \includegraphics[width=8cm, height=4cm]{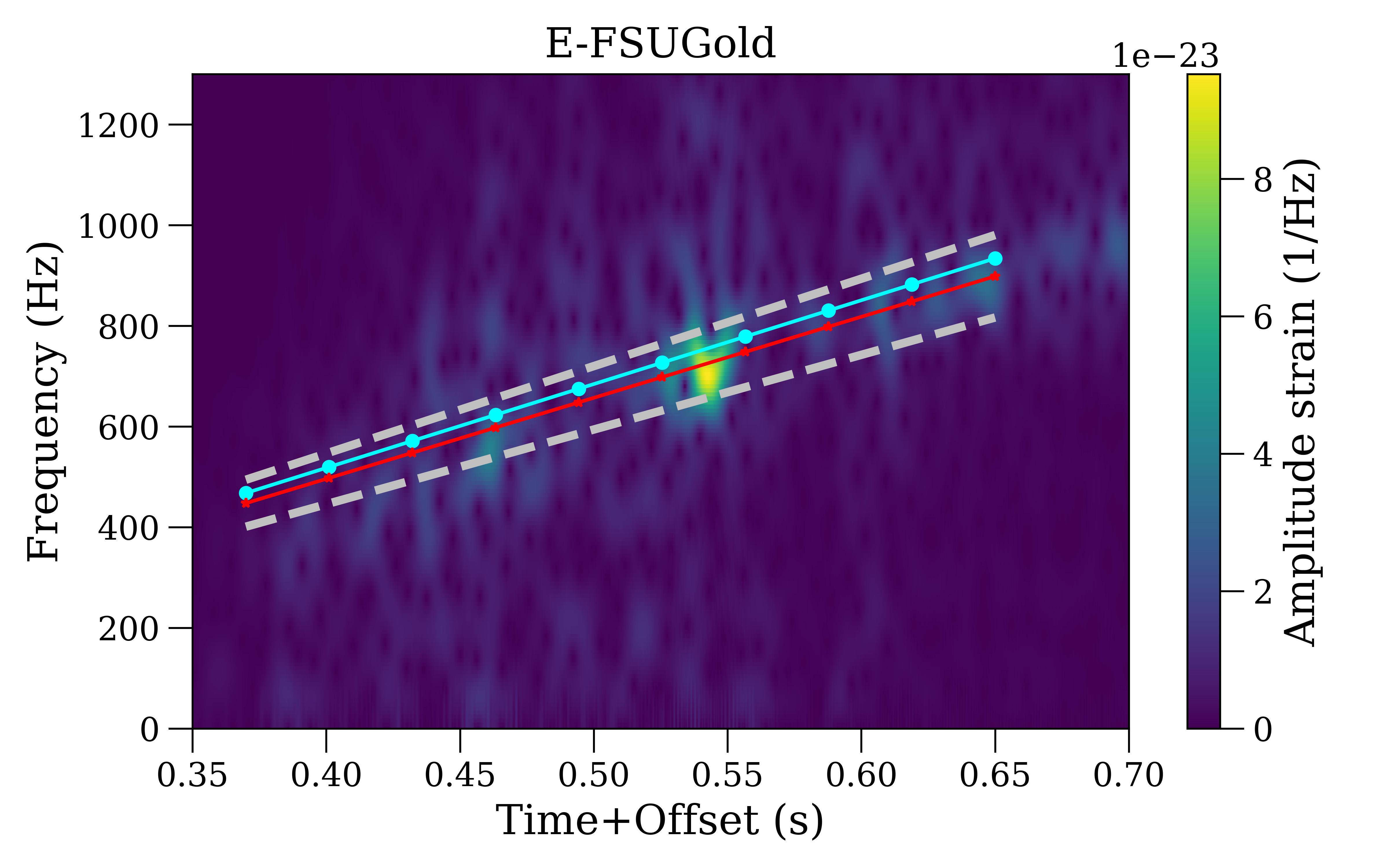}
        \hfill
            \includegraphics[width=8cm, height=4cm]{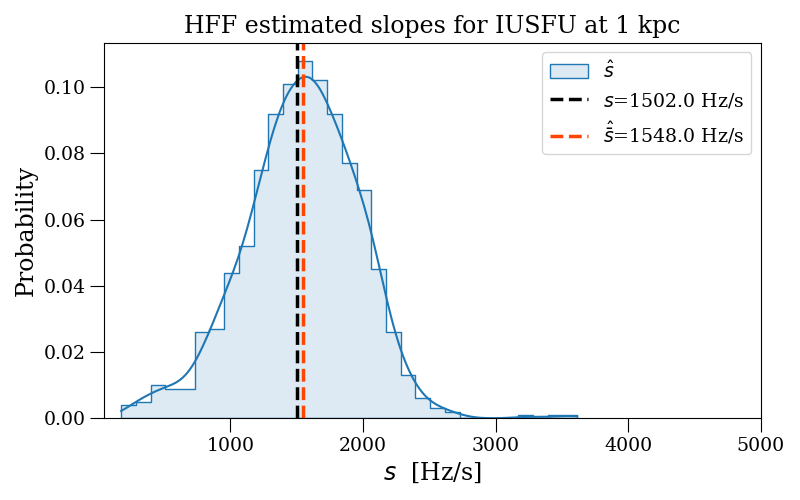}   
        \hfill
            \includegraphics[width=8cm, height=4cm]{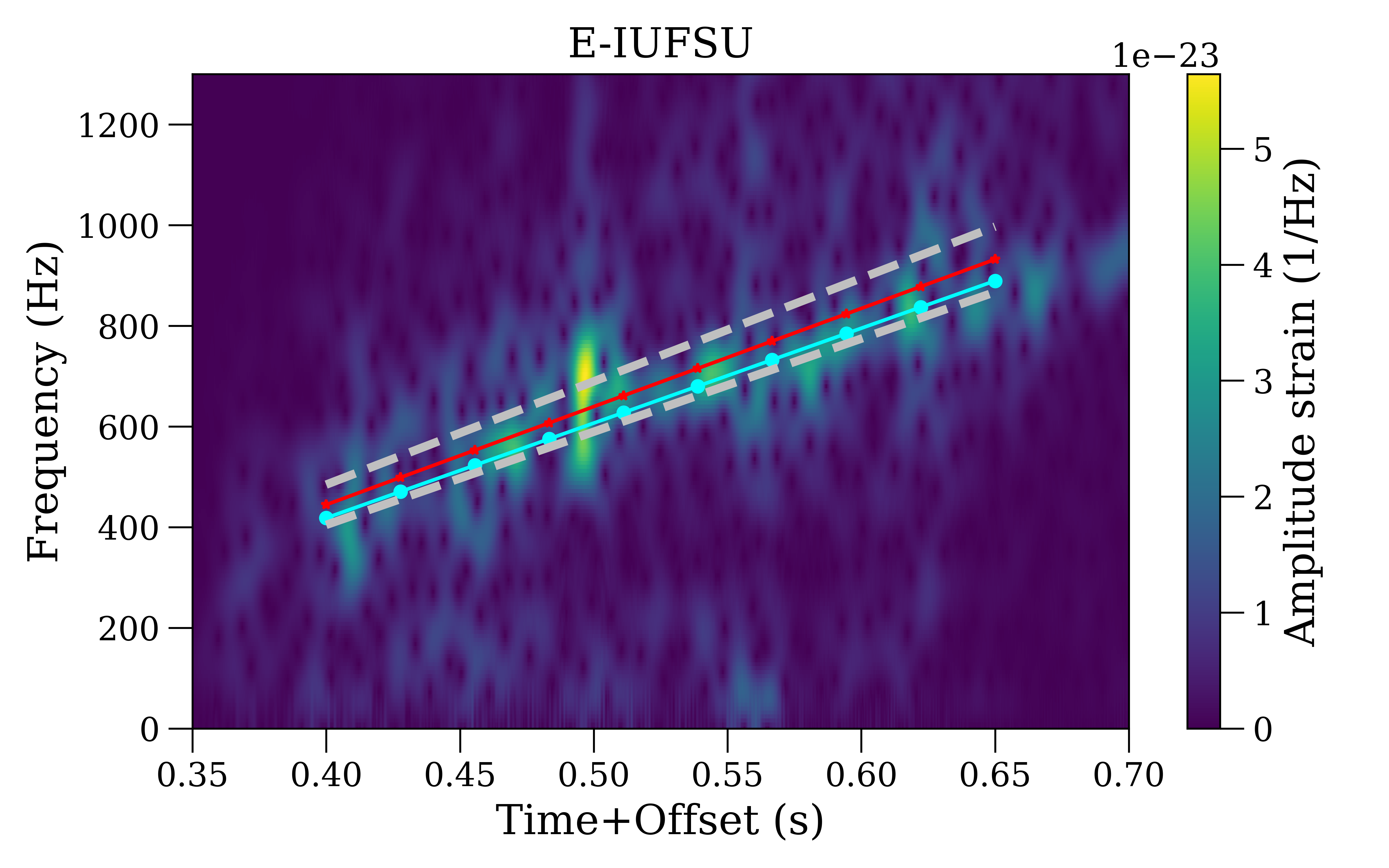}
        \hfill
            \includegraphics[width=8cm, height=4cm]{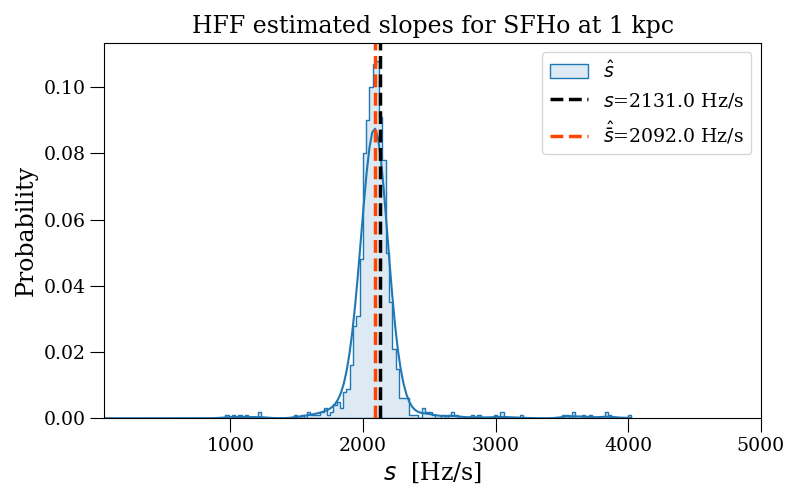}
        \hfill
            \includegraphics[width=8cm, height=4cm]{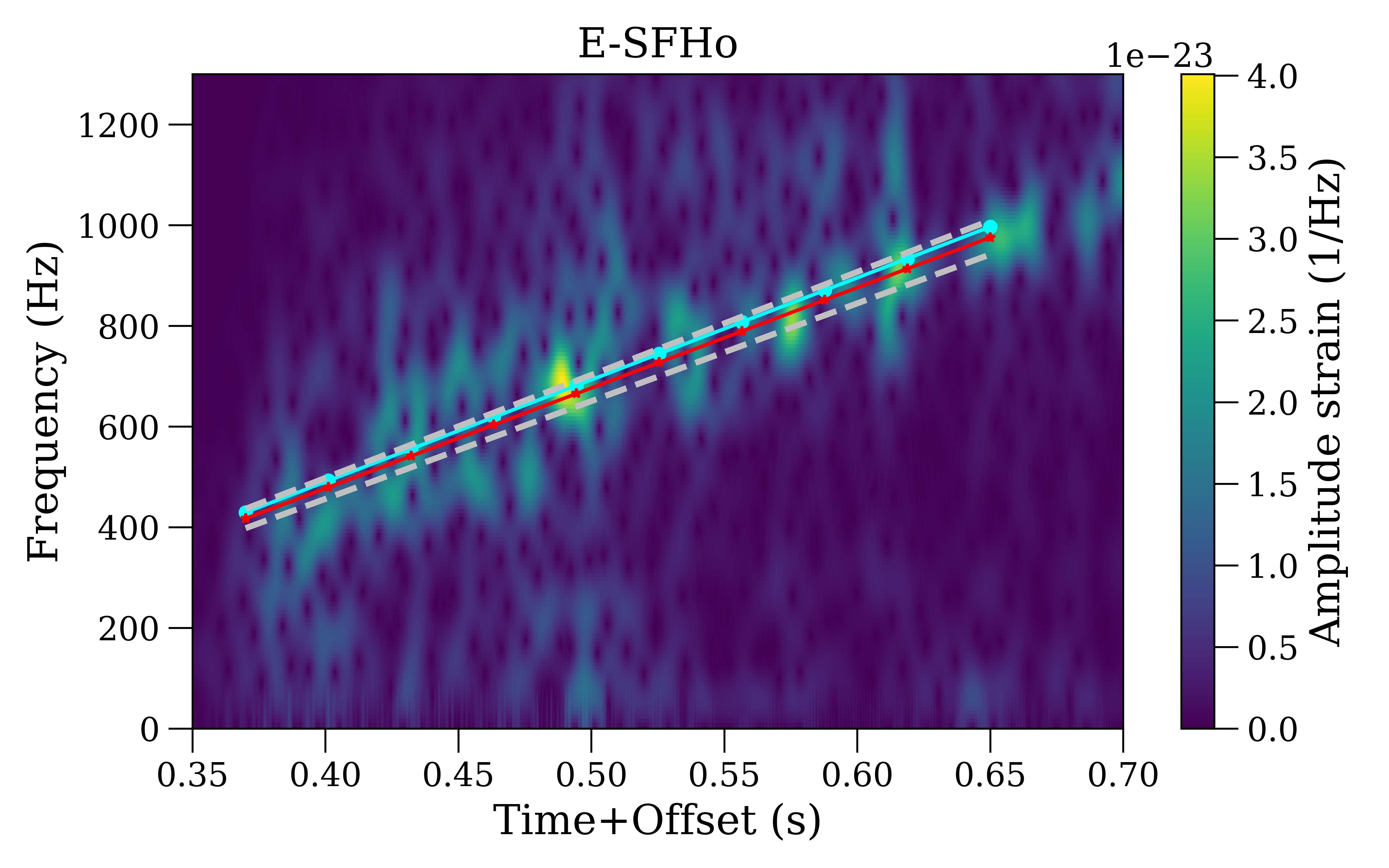}
        \hfill
            \includegraphics[width=8cm, height=4cm]{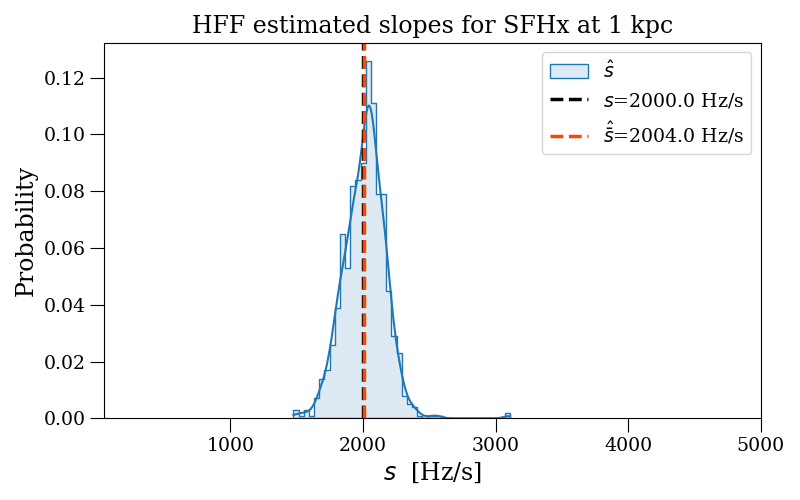}
        \hfill
            \includegraphics[width=8cm, height=4cm]{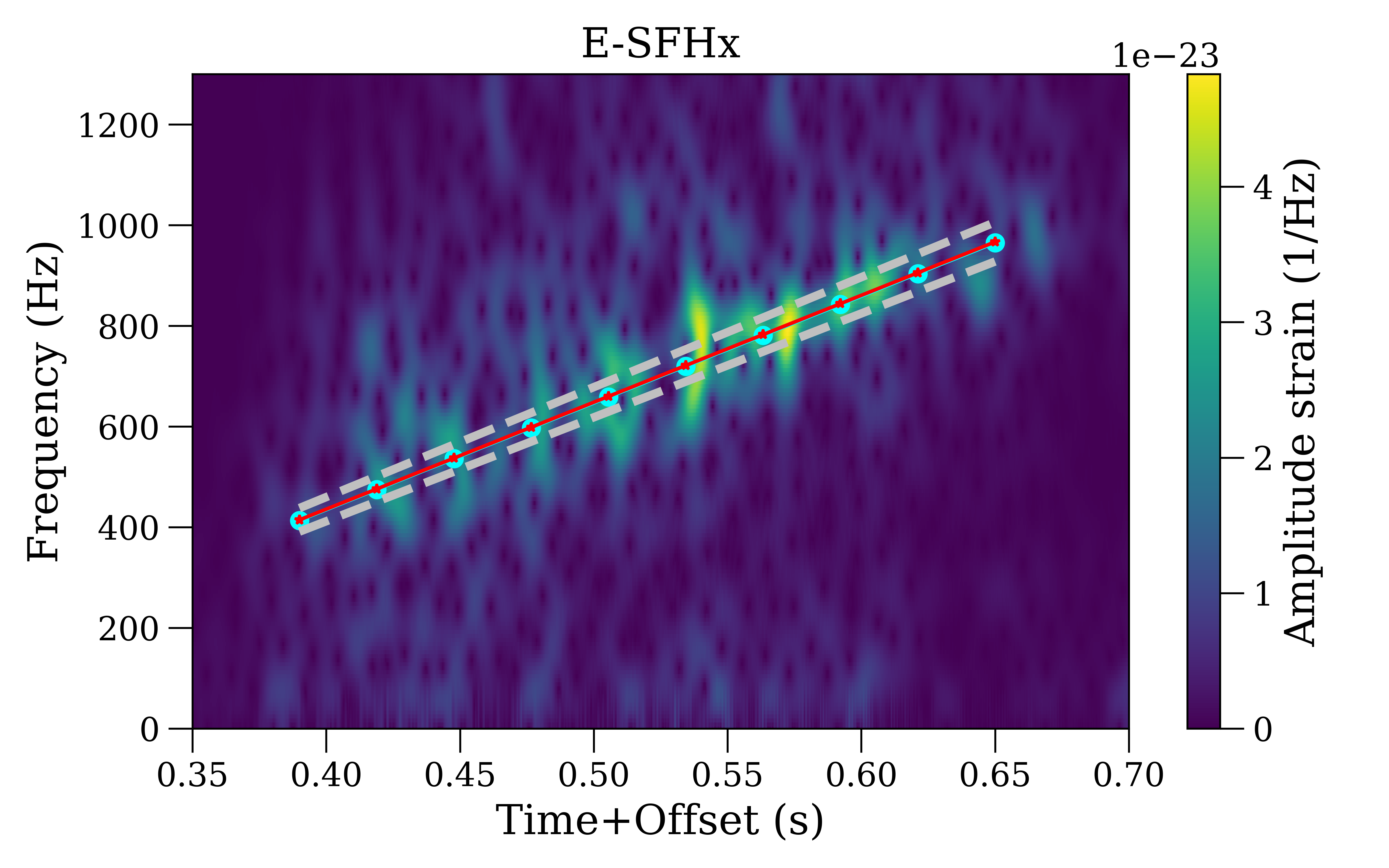}
        \caption{The histograms on the left show, in blue, the estimated slope of HFF $s$ and, with red, dashed lines the estimated means of the HFF slope $\hat{\bar{s}}$. The right column illustrates the estimation error in the E-series spectrograms. Gray, dashed lines denote the estimation errors. Solid, cyan lines with circles denote the slope of the HFF in the absence of noise. Red lines with stars correspond to the mean of the estimated HFF slope obtained through the DNN algorithm described in the text.} 
\label{fig:HFF_slopes_1kpc}
\end{figure*}

A graphical illustration of the HFF slope estimation process in real interferometric noise is shown in Figure \ref{fig:HFF_slopes_1kpc}.
In the left column, the histograms shaded in blue represent the distribution of estimated slopes in the E-series GW signals denoted as $\hat{s}$. Vertical, black, dashed lines indicate the values of the HFF slope in the absence of noise, while red, dashed lines represent the estimated means of the HFF slope, denoted as $\hat{\bar{s}}$. The right column shows the E-series spectrograms in the absence of noise with HFF approximations and estimation errors overlaid. Solid, cyan lines with circles denote HFF approximations using $s$ as the slope, and solid, red lines with stars indicate HFF approximations using $\bar{s}$ as the slope. The dashed, gray lines denote the estimation errors, in line with the standard deviations reported in Table \ref{tab:HFF_metrics} for a detection distance of 1 kpc.

\begin{table*}[!ht]
    \begin{center}
        \begin{tabular}{ l c c c c c c c c c c c c c } 
            \hline
            \hline
             & & \multicolumn{4}{c}{1 kpc} & \multicolumn{4}{c}{5 kpc} & \multicolumn{4}{c}{10 kpc} \\
             \cmidrule(lr){3-6}\cmidrule(lr){7-10}\cmidrule(lr){11-14}
            EOS & $s$ & $\hat{\bar{s}}$ & STD & RMSE & MAPE & $\hat{\bar{s}}$ & STD & RMSE & MAPE  & $\hat{\bar{s}}$ & STD & RMSE & MAPE\\
                & [Hz~s$^{-1}$] & [Hz~s$^{-1}$] & [Hz~s$^{-1}$] & [Hz~s$^{-1}$] & [\%] & [Hz~s$^{-1}$] & [Hz~s$^{-1}$] & [Hz~s$^{-1}$] & [\%] & [Hz~s$^{-1}$] & [Hz~s$^{-1}$] & [Hz~s$^{-1}$] & [\%]\\
            \hline
            DD2 & 1398 & 1402 & 103.54 & 301.72 & 5.6 & 1742 & 293.13 & 223.11 & 21 & 2377 & 765.30 & 635.93 & 32\\ 

            FSUGold & 1665 & 1610 & 126.93 & 228.84 & 1.2 & 2002 & 258.56 & 332.91 & 15 & 2544 & 678.70 & 468.19 & 27\\ 

            IUFSU & 1502 & 1566 & 100.66 & 229.11 & 3.1 & 1812 & 201.69 & 154.25 & 17 & 2404 & 555.41 & 526.20 & 24\\ 

            SFHo  & 2131 & 2092 & 52.63 & 249.39 & 0.5 & 2501 & 313.45 & 256.74 & 18 & 2670 & 401.92 & 486.82 & 21\\ 

            SFHx & 2000 & 2003 & 60.43 & 159.30 & 0.6 & 2311 & 298.12 & 106.10 & 12 & 2687 & 368.12 & 416.60 & 20\\ 
            \hline
        \end{tabular}
        \caption{HFF slope estimation results in real interferometric noise for distances of 1 kpc, 5 kpc, and 10 kpc. The slope estimation without noise is reported for each E-series signal in the second column. Then, for each detection distance, the mean estimated slope $\hat{\bar{s}}$ is reported with its corresponding standard deviation. The root-mean-square error is given by $RMSE=\sqrt{(\sum_i(s_i-\bar{s}_i)/n)}$, as a measure between the approximation using the slope without noise and the approximation using the mean slope from the DNN. Likewise, the MAPE uses the same formula shown in Table \ref{tab:comp}, but the error is calculated between the approximation using the slope determined without noise and the approximation using the mean slope from the DNN.}
        \label{tab:HFF_metrics}
    \end{center}
\end{table*}

Table \ref{tab:HFF_metrics} shows the results of using the DNN algorithm for the E-series GW signals at distances of 1, 5, and 10 kpc. In each case, a mean slope $\hat{\bar{s}}$ is found using the DNN, and the RMSE and MAPE values relate the HFF approximation using $\hat{\bar{s}}$ to the value of the slope without noise $s$. Table \ref{tab:HFF_intercept} reports the same results for the HFF starting frequency, denoted as $\beta$ in Section \ref{subsec:HFF_lin}.
\begin{table*}[!ht]
    \begin{center}
        \begin{tabular}{ l c c c c c c c c c c c c} 
            \hline
            \hline
             & \multicolumn{4}{c}{1 kpc} & \multicolumn{4}{c}{5 kpc} & \multicolumn{4}{c}{10 kpc} \\
             \cmidrule(lr){2-5}\cmidrule(lr){6-9}\cmidrule(lr){10-13}
            EOS & $\bar{\beta}$ & STD & RMSE & MAPE & $\bar{\beta}$ & STD & RMSE & MAPE & $\bar{\beta}$ & STD & RMSE & MAPE\\
                & [Hz] & [Hz] & [Hz] & [\%] & [Hz] & [Hz] & [Hz] & [\%] & [Hz] & [Hz] & [Hz] & [\%]\\
            \hline
            DD2 & 419 & 5.21 & 11.23 & 4.32 & 531 & 12.94 & 13.54 & 10.18 & 649 & 27.39 & 31.30 & 23.62\\ 

            FSUGold & 429 & 8.51 & 13.47 & 6.05 & 550 & 18.92 & 27.14 & 15.73 & 714 & 31.32 & 34.01 & 30.98\\ 

            IUFSU & 378 & 9.01 & 14.04 & 6.67 & 505 & 20.04 & 29.37 & 16.34 & 708 & 33.51 & 36.37 & 33.75\\ 

            SFHo  & 364 & 8.39 & 12.57 & 6.01 & 481 & 19.49 & 27.01 & 15.24 & 698 & 30.21 & 33.23 & 33.13\\ 

            SFHx & 408 & 5.87 & 11.74 & 4.94 & 523 & 15.50 & 25.21 & 11.36 & 660 & 29.94 & 32.12 & 26.17\\ 
            \hline
        \end{tabular}
        \caption{HFF starting frequency estimation results, in real interferometric noise, for distances of 1 kpc, 5 kpc, and 10 kpc. $\bar{\beta}$ is the mean HFF starting frequency obtained from the DNN, and it is reported with its corresponding standard deviation for each Galactic distance. The RMSE and MAPE are computed using the same formulas described in Table \ref{tab:HFF_metrics} and Table \ref{tab:comp}, respectively. These quantities are computed with respect to their values as determined without noise.}
        \label{tab:HFF_intercept}
    \end{center}
\end{table*}

\begin{figure*}[!ht]
        \centering
            \includegraphics[width=8.5cm]{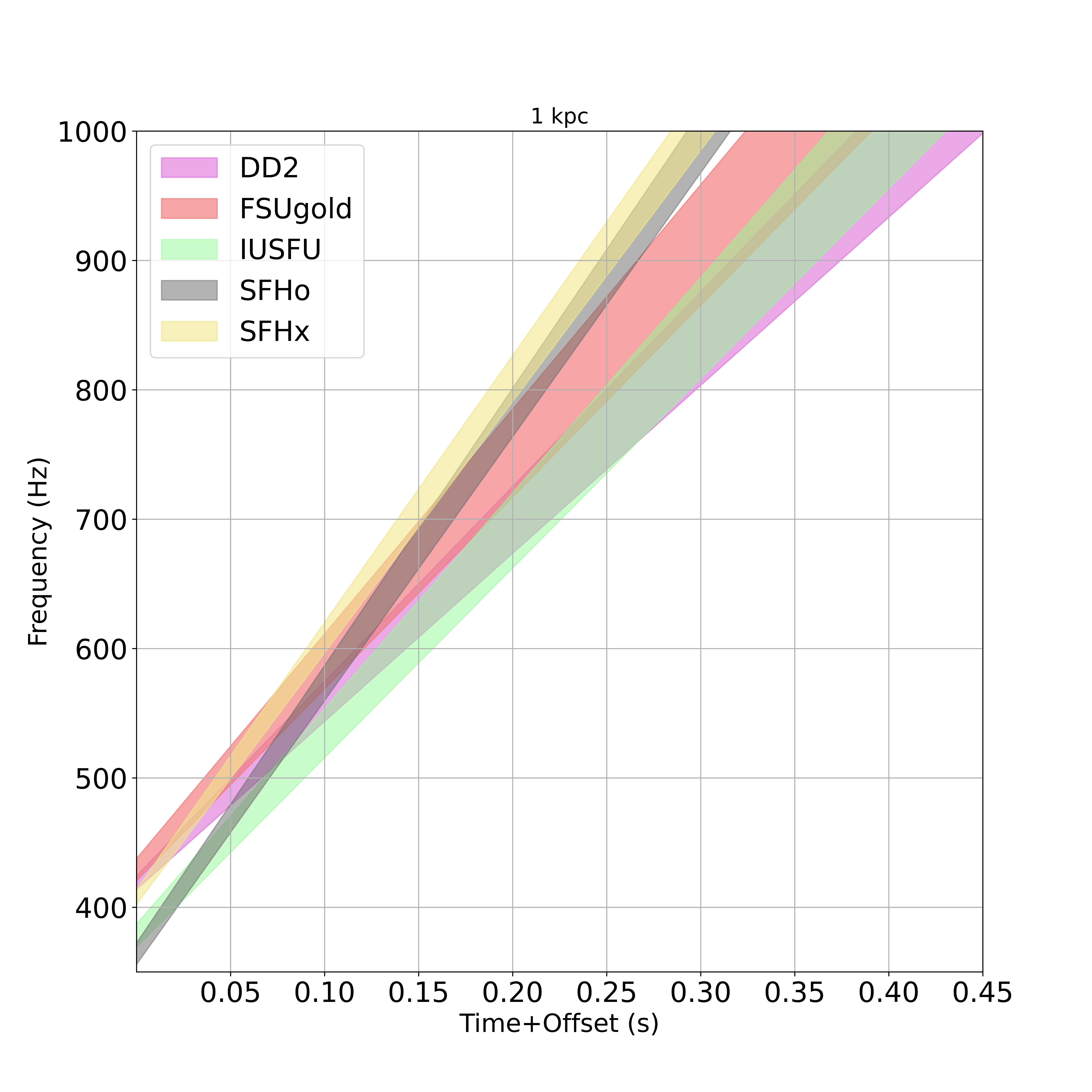}
        \hfill
            \includegraphics[width=8.5cm]{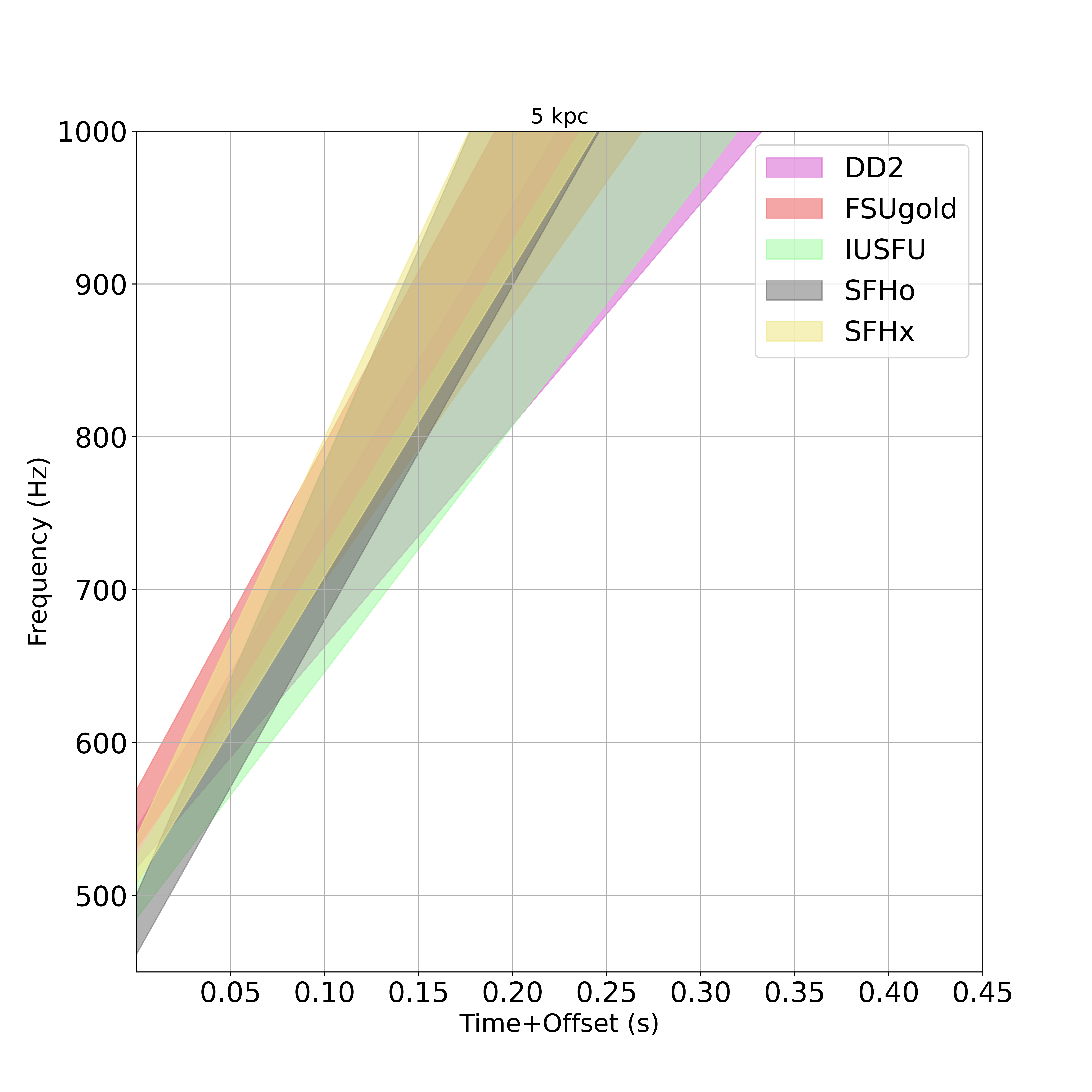}
        \hfill
            \includegraphics[width=8.5cm]{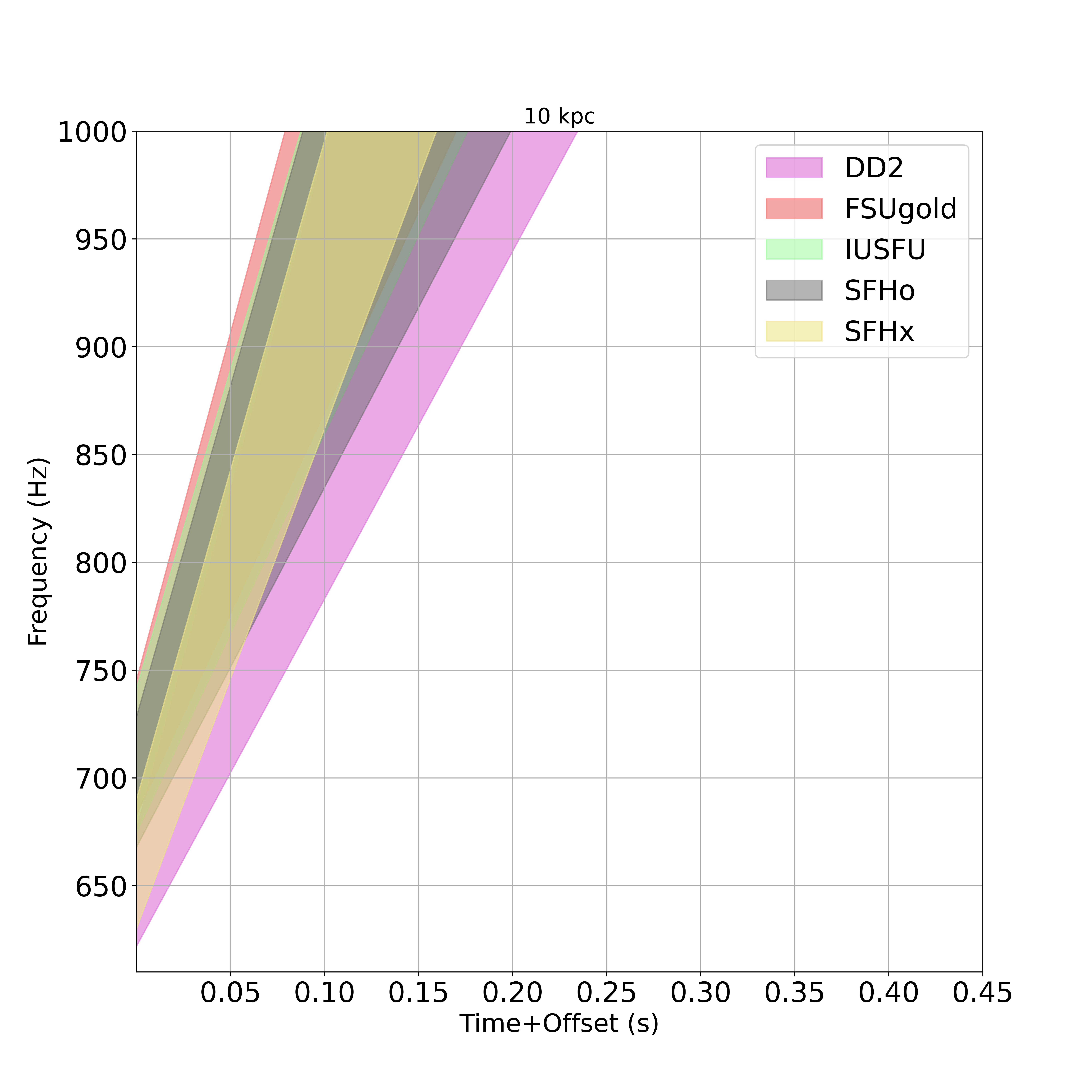}

        \caption{HFF slopes estimated by our DNN model including the range of variability of the HFF starting frequency across E-series models for detection distances of 1 kpc, 5 kpc, and 10 kpc. The figures visually represent the estimation errors in both HFF slope and starting frequency. The color band surrounding the data points corresponds to the STD around the mean of the estimated values of $\hat{\bar{s}}$ and $\bar{\beta}$ as is reported in table \ref{tab:HFF_metrics} and \ref{tab:HFF_intercept}.}
\label{fig:HFF_est_error}
\end{figure*}

Figure \ref{fig:HFF_est_error} shows the linear estimation of the HFF for the E-series GW signals with the $\pm1\sigma$ results shaded for each model for each Galactic detection distance. This shading encompasses the standard deviations in both HFF slope and HFF starting frequency. As seen in Table \ref{tab:HFF_intercept}, the mean HFF starting frequency varies with detection distance. This is reflected in Figure \ref{fig:HFF_est_error}, as each plot has a different HFF starting frequency on the frequency axis.

These results all used the cWB parameters described at the end of Section \ref{subsec:cWB}, which are optimized for detection of GW signals. In the case of a detection already being confirmed, either through neutrino detection or through comparison with other GW detectors, it is possible that the cWB parameters could be changed to optimize analysis of the HFF rather than detection. One such parameter is the BPP, a dimensionless parameter that controls the reconstruction of the CCSN GW signal in the wavelet decomposition. We initially set the BPP to 0.05, and then increased it to 0.1 to gain a deeper understanding of the interferometric noise structure and its correlation with the HFF slope estimation. Table \ref{tab:HFF_metrics_1kpc_BPP_0.1} shows how the estimation of the HFF slope varies with the BPP at different Galactic distances. With increasing BPP, the evidence indicates that the STD associated with each EOS tends to decline. This is indicated by the $STD_{\rm improvement}$ value. This implies that it is possible to investigate the properties of the interferometric noise to extend the physical information provided by these events.

\begin{table*}[!ht]
    \begin{center}
        \begin{tabular}{ l c c c c c c c } 
            \hline
            \hline
            EOS & $s$ & $\hat{\bar{s}}$ & STD & STD improvement & MAPE\\
                & [Hz~s$^{-1}$]    &   [Hz~s$^{-1}$]  &   [Hz~s$^{-1}$]    &   [\%]  & [\%]   \\
            \hline
            DD2 & 1398 & 1400 & 82.83 & 20 & 4.2\\ 

            FSUGold & 1665 & 1615 & 95.19 & 25 & 0.7\\ 

            IUFSU & 1502 & 1560 & 80.66 & 20 & 2.6\\ 

            SFHo  & 2131 & 2096 & 49.65 & 25 & 0.3\\ 

            SFHx & 2000 & 2003 & 40.23 & 22 & 0.2\\ 
        \hline
        \end{tabular}
        \caption{The results of the HFF slope estimation are presented, with a variation in the cWB BPP parameter operation of BPP=0.1. Each column contains the values associated with the slope $s$, the slope estimated using the DNN $\hat{\bar{s}}$, the standard deviation STD, the improvement in STD compared to the BPP=0.05 used in this study, and the MAPE. These results are for a detection distance of 1 kpc.}
          \label{tab:HFF_metrics_1kpc_BPP_0.1}
    \end{center}
\end{table*}

\section{Discussion}\label{Sec:Disc}

From Figure \ref{fig:HFF_slopes}, it is clear that in the absence of noise the slope of the HFF of GWs generated from CCSN simulation is distinguishable for each EOS used. Our analysis in the absence of noise shows a variation on the order of 50\% in the HFF slope for the E-DD2 simulation versus the E-SFHo simulation. These represent the lowest and highest slope values of the HFF, respectively. The smallest variation is between E-SFHo and E-SFHx, varying by 10\% in HFF slope. This highlights the level of accuracy that is needed to detect these differences. When noise is introduced, a slope error estimation of up to 50\% has the possibility to discern differences in EOS slopes in the extreme cases. Likewise, a slope error estimation below 10\% has reached the necessary level of accuracy to differentiate the smallest differences in slope we observe in the noiseless case.

Using the methods established in \cite{PhysRevD.108.084027}, we show the capability of using the slope of the HFF to investigate CCSNe. We note that the definition of the beginning of the HFF signal is less relevant when the signal is injected with real LIGO noise because, in this case, strains below noise levels cannot be distinguished. Thus, the HFF signal starts, in the detections with real noise, where the strain overcomes the noise threshold of the detector. The mean HFF starting frequency reported in Table \ref{tab:HFF_intercept} is then a result of the range of starting frequencies that may be detected depending on when the signal overcomes the noise threshold.

Figure \ref{fig:HFF_est_error} shows the estimation errors for both HFF slope and HFF starting frequency for 1, 5, and 10 kpc. At 1 kpc, there are regions with no overlap in estimation error between each E-series model. We see the emergence of two groupings: the steep-sloped E-SFHo and E-SFHx models and the more-shallow-sloped E-DD2 and E-IUFSU models. The estimation error of E-FSUGold crosses both groupings, having a high HFF starting frequency but more shallow slope. Having low estimation error in the HFF starting frequency allows for clear distinctions between models with similar slopes. E-SFHo and SFHx are similar in slope, but well resolved and separated in HFF starting frequency. The same is true for E-IUFSU and E-DD2. The minimum slope resolvability is between E-SFHo and E-SFHx, which differ by 4.4\% and have HFF starting frequencies differing by 12\%. The minimum HFF starting frequency resolvability is between E-DD2 and E-FSUGold, which differ by 2.4\% and have HFF slopes differing by 15\%. In other words, for EOS models with larger differences in HFF starting frequency we can resolve smaller differences in HFF slope. Conversely, for EOS models with roughly similar HFF starting frequencies we require a greater separation in HFF slope in order to resolve them.
 
At 5 kpc, the HFF starting frequency is notably higher across all models. This is due to the weaker signal strength at the beginning of the HFF when compared to the increased noise, so the HFF is detected at a later time in the signal. We thus lose the clear period of resolvability at early signal times, as seen in the 1 kpc detection. Nonetheless, E-SFHx and E-SFHo remain largely distinguishable from E-DD2 and E-IUFSU, though distinguishing within these groups is more difficult. E-FSUGold remains distinguishable at early times, but not at later times.

At 10 kpc, as the the reconstruction of the earlier HFF signal is further affected by the noise, the HFF starting frequency continues to rise. The resolvability between the groups, seen at 1 and 5 kpc, is lost. However, the upper bounds of E-FSUGOld and lower bounds of E-DD2 in both HFF starting frequency and slope are still resolvable. With the expected improvements in sensitivity in future runs, when interferometers will reach design sensitivity and when the number of interferometers will increase, we expect the uncertainty in the starting frequency to decrease, as well as the uncertainty in the slope. For this reason the results presented here should be thought of as conservative.

These results are consistent with the noise data from the LIGO O3 run conducted in 2019 through 2020. The LIGO O4 run is currently underway, with improved sensitivity, but O4 noise data is not yet available. With the improved sensitivity of future detectors, we expect the ability to measure the slope of the HFF below 1 kHz to improve as well. In our models, SNR is decreased with distance to the event due to the inverse relationship of GW signal amplitude with distance. This means that improved SNR for future detectors will effectively shift our results to larger distances by the same factor. In the case of the order-of-magnitude improvements to detection sensitivity currently projected for the Einstein Telescope \cite{ET} and Cosmic Explorer \cite{Cosmic_Explorer}, our 1 kpc results would be, conservatively, representative of 10 kpc detections.

As noted in Section \ref{sec:2d3d}, the strain amplitudes of GW signals from two-dimensional CCSN simulations are overestimated in comparison to those from three-dimensional CCSN simulations. This means that the GW signals considered in this study are likely stronger than the expected GW signal from a Galactic CCSN. We noted that the power spectral density is $\sim 8.4$ times higher for the two-dimensional model than for the three-dimensional model, with all other model parameters being identical. Due to the inverse linear relationship with signal strength and distance, this means we can approximate the two-dimensional results as three-dimensional results that are $\sim 8.4$ times closer than the signals analyzed in Figure \ref{fig:HFF_est_error}.

\section{Conclusion}\label{Sec:conc}

This paper shows the capability of using HFF slope measurements to distinguish between the possible EOS at play in CCSN events within 1 kpc for current-generation detectors. The improved sensitivity of future detectors will improve our ability to distinguish between different EOS, for distances at least an order of magnitude larger. We also expect further benefits from customized tuning of cWB for PE purposes, as shown in  \citet{Lin_2023}. These improved sensitivities will also allow for an estimation of the nonlinear time dependence of the HFF, given the signal above 1 kHz will become detectable. We leave a study of this phase of the evolution of the HFF to future work. 

In this study, we investigated the effects on the HFF of varying the EOS in isolation, without varying progenitor parameters. Not only do parameters like rotation rate \cite{Andresen_2019, Pajkos_2019, Pajkos_2021, Pan_2021, Takiwaki_2021} and mass \cite{Andresen_2017, Vartanyan_2019, Mezz_D_GW_2023} affect the slope of the HFF, but parameters varied simultaneously affect the slope in more complex ways. As seen in \citet{Wang_2024}, varying the internal structure at the same time as rotation rate leads to HFF slopes that are decreased less than when there is no internal structure variation. Similarly, \citet{Jardine_2022} show that the presence of strong magnetic fields may also counteract the effect of rotation on the slope of the HFF.

Future studies investigating the combined effects of varying intrinsic properties of the progenitor and the EOS are necessary in order to develop a complete understanding of what can be learned about the EOS from the detection of the HFF in the next Galactic CCSN. The study conducted here can be viewed as providing limits on what we can hope to learn.

\section*{Acknowledgements}
This research used resources of the Oak Ridge Leadership Computing Facility at the Oak Ridge National Laboratory, which is supported by the Office of Science of the U.S. Department of Energy under Contract No. DE-AC05-00OR22725. ACL acknowledges a CONACYT scholarship and CONACyT Network Project No. 376127 {\it Sombras, lentes y ondas gravitatorias generadas por objetos compactos astrofísicos.} A.M. acknowledges support from the National Science Foundation's Gravitational Physics Theory Program through grants PHY-1806692 and PHY 2110177. M.Z. is supported by the National Science Foundation Gravitational Physics Experimental and Data Analysis Program through award PHY-2110555. P.M. is supported by the National Science Foundation through its employee IR/D program.

This research has made use of data or software obtained from the Gravitational Wave Open Science Center (\href{https://gwosc.org}{gwosc.org}), a service of the LIGO Scientific Collaboration, the Virgo Collaboration, and KAGRA. This material is based upon work supported by NSF's LIGO Laboratory, which is a major facility fully funded by the National Science Foundation, as well as the Science and Technology Facilities Council (STFC) of the United Kingdom, the Max-Planck-Society (MPS), and the State of Niedersachsen/Germany for support of the construction of Advanced LIGO and construction and operation of the GEO600 detector. Additional support for Advanced LIGO was provided by the Australian Research Council.
Virgo is funded, through the European Gravitational Observatory (EGO), by the French Centre National de Recherche Scientifique (CNRS), the Italian Istituto Nazionale di Fisica Nucleare (INFN) and the Dutch Nikhef, with contributions by institutions from Belgium, Germany, Greece, Hungary, Ireland, Japan, Monaco, Poland, Portugal, and Spain. KAGRA is supported by the Ministry of Education, Culture, Sports, Science, and Technology (MEXT), the Japan Society for the Promotion of Science (JSPS) in Japan, the National Research Foundation (NRF) and Ministry of Science and ICT (MSIT) in Korea, and the Academia Sinica (AS) and National Science and Technology Council (NSTC) in Taiwan.

\appendix
\section{Curvature of the HFF}\label{app:Curv}

As mentioned in Section \ref{sec:GW_A}, a natural extension to this analysis would be to examine the curvature of the HFF. Section \ref{subsec:L} showed that the current detection probability of the HFF above 1 kHz is low, and this was part of our justification for examining only the first-order time dependence of the HFF. Here, we show that approximating the HFF time dependence to second-order time results in different estimates of curvature for the HFF signal below 1 kHz, and the curvature for the signal up to 2 kHz.

\begin{figure*}[!ht]
        \centering
            \includegraphics[width=8.6cm]{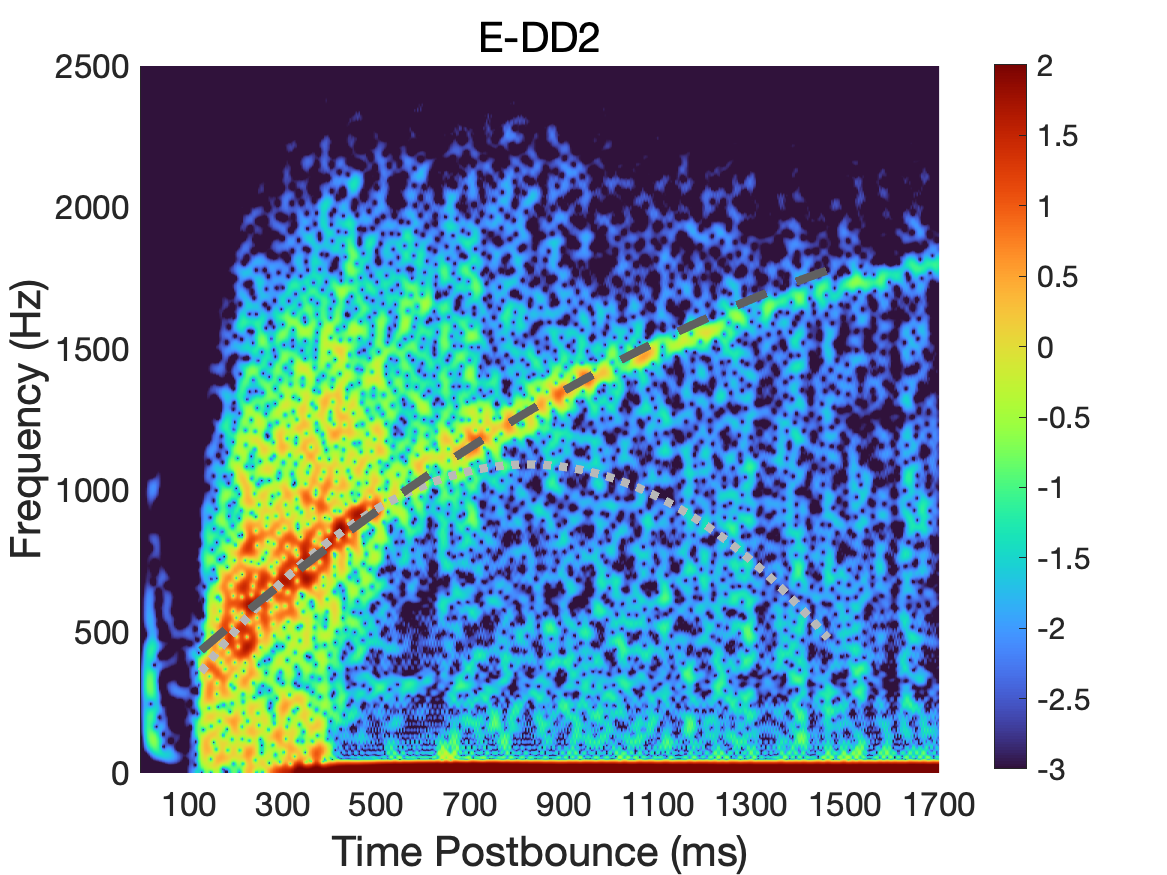}
            \includegraphics[width=8.6cm]{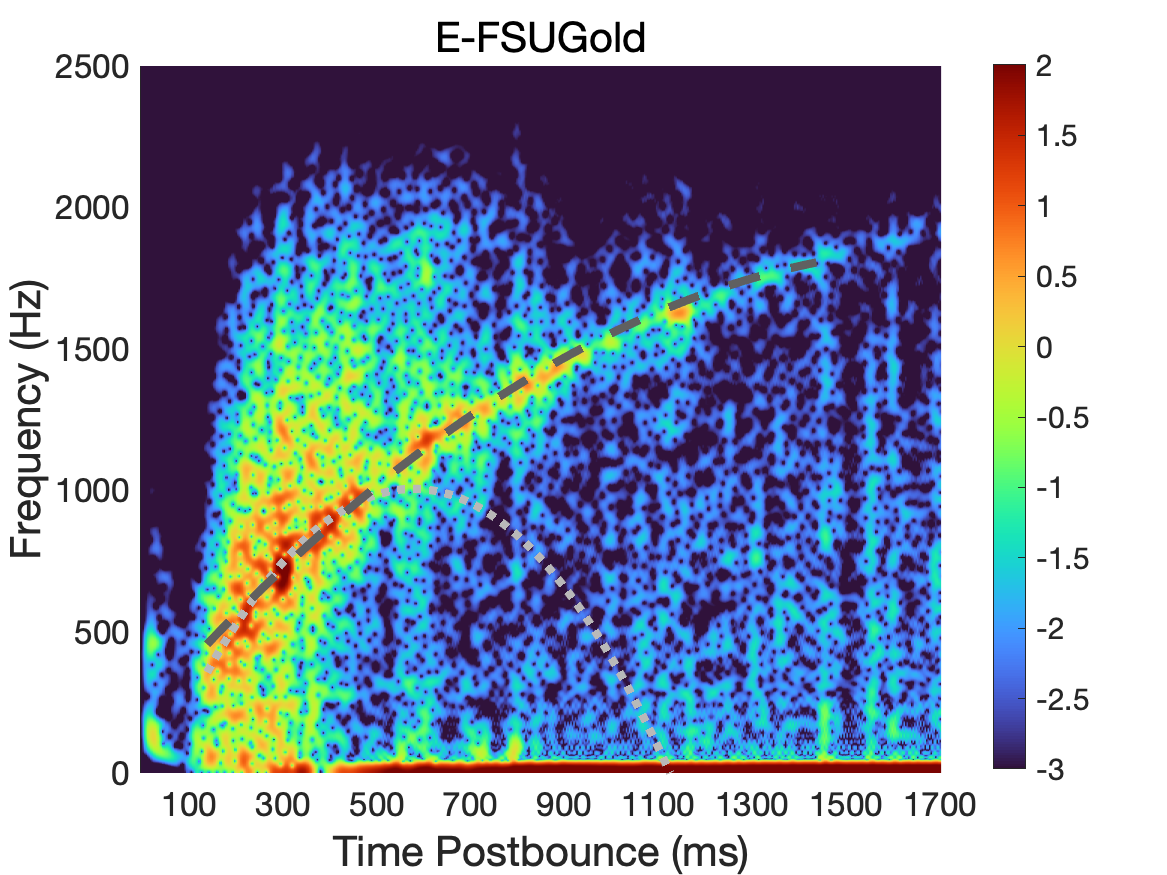}
        \hfill
            \includegraphics[width=8.6cm]{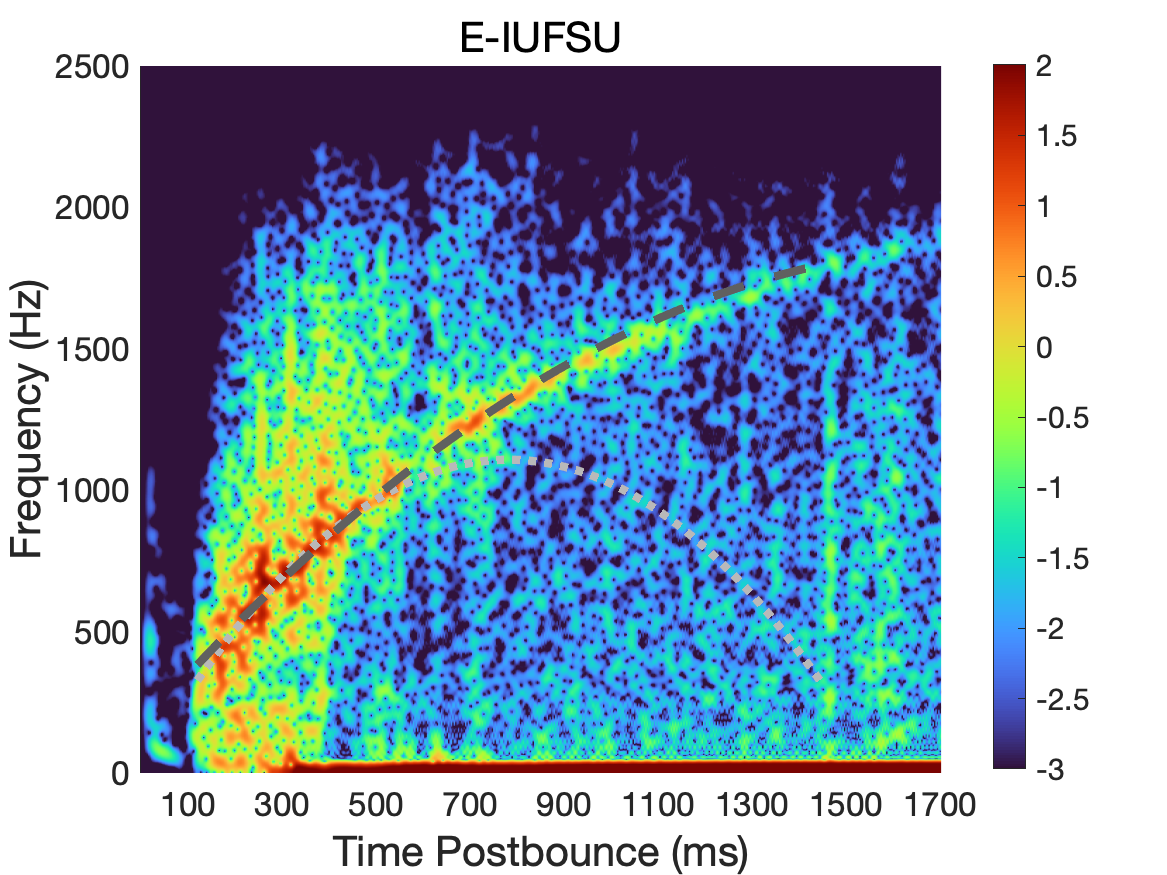}
            \includegraphics[width=8.6cm]{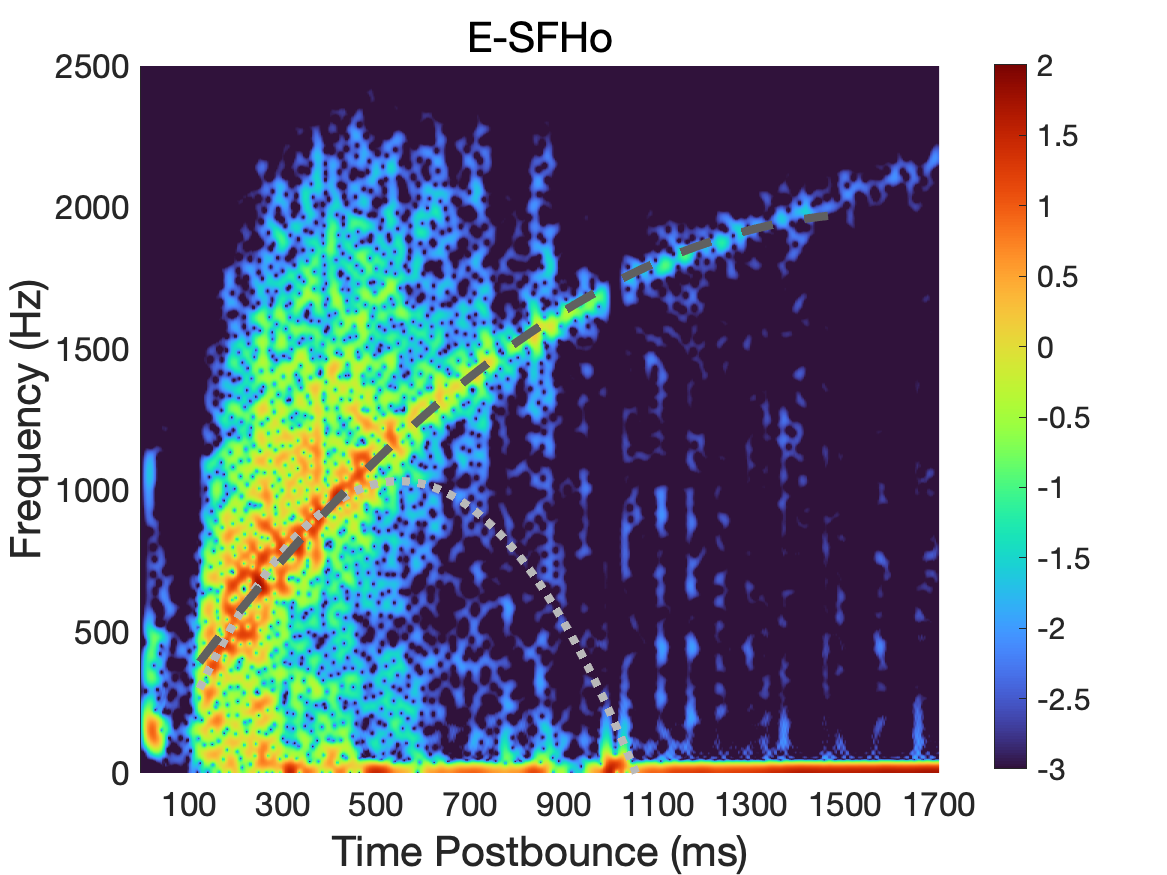}
        \hfill
            \includegraphics[width=8.6cm]{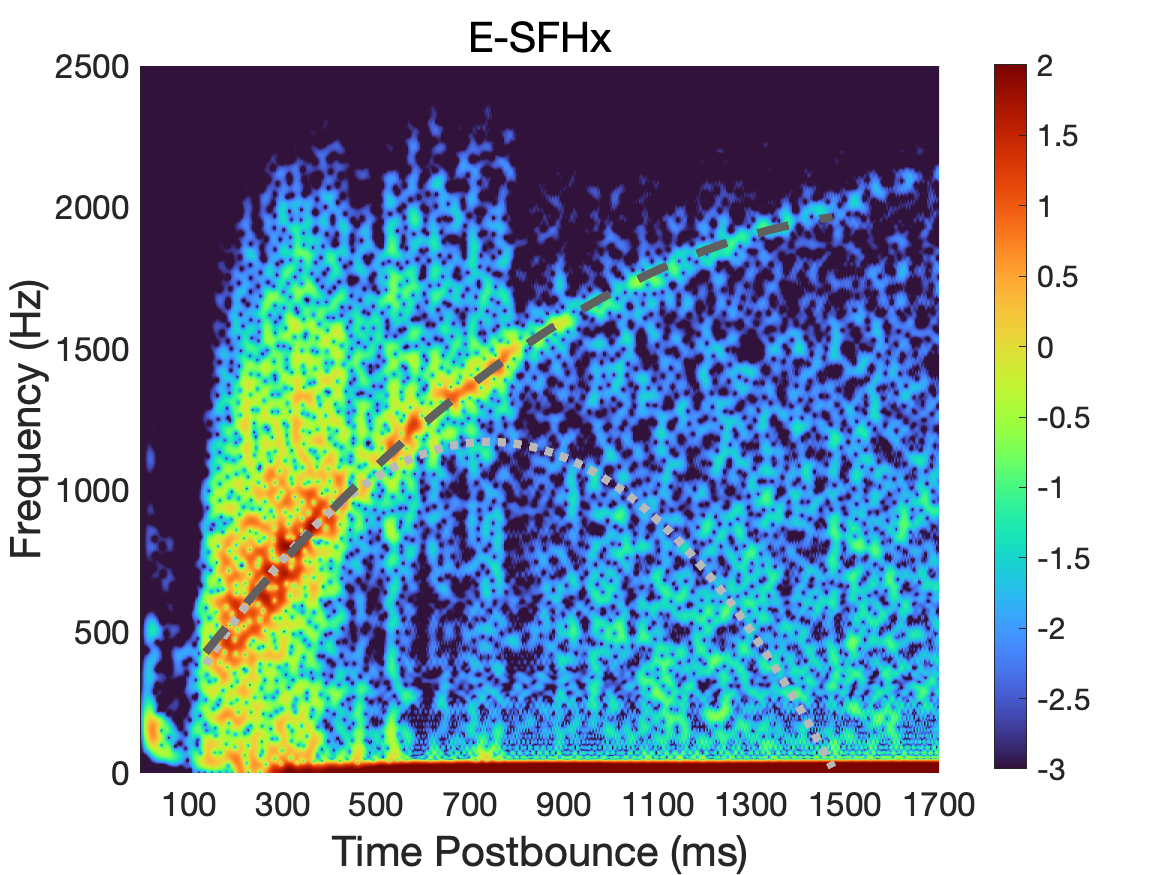}
        \caption{Spectrograms for each EOS, with the color axis showing the logarithm of the power spectrum, $\log_{10}(P)$. The light gray curve is determined using the coefficients of the second-order regression of the HFF signal below 1 kHz, while the dark gray line uses the coefficients determined from the regression of the HFF signal below 2~kHz.} 
        \label{fig:spec_O2}
\end{figure*}

The procedure for defining the starting time of the signal is the same as described in Section \ref{subsec:HFF_lin}. Instead of performing a first-order regression, we now allow our regression to include terms up to second order in time. In Figure \ref{fig:spec_O2}, the light, gray, dotted curve uses the coefficients determined for the HFF signal below 1 kHz. The dark, gray, dashed curve uses coefficients determined for the HFF signal below 2 kHz. Table \ref{tab:O2} shows the coefficients for each approximation. 

 \begin{table*}[!ht]
    \begin{center}
        \begin{tabular}{ l c c c c c c } 
            \hline
            \hline
              &\multicolumn{3}{c}{HFF Below 1 kHz} & \multicolumn{3}{c}{HFF Below 2 kHz}\\
             \cmidrule(lr){2-4}\cmidrule(lr){5-7}
            EOS & $\gamma$ & $\alpha$ & $\beta$ & $\gamma$ & $\alpha$ & $\beta$ \\
                & [Hz~s$^{-2}$] & [Hz~s$^{-1}$] & [Hz] & [Hz~s$^{-2}$]  & [Hz~s$^{-1}$] & [Hz] \\
            \hline
            DD2 & -1504 & 2100 & 357.4 & -321.5 & 1436 & 433.0 \\ 
            
            FSUGold & -3354 & 2951 & 355.8 & -541.3 & 1743 & 454.9 \\ 
            
            IUFSU & -1776 & 2355 & 326.9 & -524.9 & 1762 & 382.3\\ 
            
            SFHo  & -4047 & 3448 & 297.6 & -745.7 & 2177 & 391.6\\ 
        
            SFHx & -2148 & 2601 & 383.6 & -673.3 & 2052 & 426.0\\
            \hline
        \end{tabular}
        \caption{Coefficients for the second-order approximation in time of the HFF below 1 kHz and below 2 kHz. Each column represents a coefficient in the approximation of the HFF time dependence of the form $f(t)=\gamma t^2 + \alpha t +\beta$.}
          \label{tab:O2}
    \end{center}
    \end{table*}

We see that the curve using the $<1$~kHz HFF coefficients overestimates the curvature when compared to the curve that uses the $<2$~kHz coefficients. Further, the first-order coefficients reported in Table~\ref{tab:O2} for the $<2$~kHz HFF differ from the first-order coefficients for the linear approximation given in Table~\ref{tab:comp} by $<3.3$\% for all models except E-IUFSU, which differs by 8\%. This shows that our measurement of the first-order time dependence of the HFF in the absence of noise below 1 kHz is in agreement with the first order time dependence of the HFF below 2 kHz when time dependence up to second order is included in our regression model. 

\bibliography{pr_add_journals,Bib}
\end{document}